\documentclass[twocolumn,bibyear]{aa}

\usepackage[varg]{txfonts}

\usepackage{xcolor}

\newcommand{\nc}{\newcommand}

\nc{\Msun}{\ensuremath{\mathrm{M}_\odot}}
\nc{\Rsun}{\ensuremath{\mathrm{R}_\odot}}
\nc{\cmcub}{\mbox{cm$^{-3}$}}
\newcommand\arcdeg{\mbox{$^\circ$}}%
\newcommand{\CII}{[C \,{\sc ii}]}

\newcommand{\NII}{[N \,{\sc ii}]}
\newcommand{\CI}{[C \,{\sc i}]}

\nc\mic{\mbox{$\mu$m}}
\nc{\thCO}{$^{13}$CO}
\nc{\hcoplus}{HCO$^+$}
\newcommand{\OI}{[O \,{\sc i}]}

\nc{\cmsq}{\mbox{cm$^{-2}$}}
\nc{\kms}{\mbox{km\,s$^{-1}$}}
\nc{\lsun}{\ensuremath{\mathrm{L}_\odot}}

\nc{\CeiO}{C$^{18}$O}
\nc{\tex}{T$_{\rm ex}$}
\nc{\tkin}{T$_{\rm kin}$}
\nc{\rsun}{\ensuremath{\mathrm{R}_\odot}}
\nc{\arcs}{\ensuremath{^{\prime\prime}}}
\nc{\arcm}{\ensuremath{^{\prime}}}
\nc{\arcd}{\ensuremath{^{\circle}}}

\usepackage{graphicx}
\usepackage[varg]{txfonts}
\usepackage{natbib}
\usepackage[colorlinks=true,     linkcolor=blue, citecolor=blue, filecolor=blue, urlcolor=blue]{hyperref}

\begin{document} 

\title{Water abundance and ortho-para ratio in the S140 photon
  dominated region}

\author{Dariusz~C.~Lis\inst{1}
  \and Sofia~Lawsky\inst{2}
  \and Emeric Bron\inst{3}
  \and Franck Le Petit\inst{3}
  \and Javier~R.~Goicoechea\inst{4}
  }

\institute{Jet Propulsion Laboratory, California Institute of
  Technology, 4800 Oak Drove Drive, Pasadena, CA 91109, USA
  \and Department of Astronomy, Columbia University,
  550 West 120th Street, New York, NY 10027, USA
  \and LUX, Observatoire de Paris, Universit\'{e} PSL, CNRS, 
  Sorbonne Universit\'{e},  92190 Meudon, France
  \and Instituto de F\'{\i}sica Fundamental (CSIC), Calle Serrano 
  121-123, 28006 Madrid, Spain
}

\date{Received 2026 May 12 / Accepted 2026 October 1} 

\abstract{We analyzed archival far-infrared to submillimeter
  \emph{Herschel} PACS, SPIRE, and HIFI spectroscopic observations of
  the S140 photon dominated region (PDR) to constrain the water vapor
  abundance and ortho-para ratio in this region. Toward the high-mass
  protostar IRS1, 52 spectral lines, including 15 water lines, are
  detected with the PACS and SPIRE instruments. PDR modeling suggests
  a UV radiation field $G_0 \simeq 1 \times 10^3$ Habing units,
  and a thermal gas pressure of $5 \times 10^8$~K\,cm$^{-3}$ in this
  region. The models strongly underestimate intensities of
  high-$J$ CO lines, which may indicate the presence of hot
  shocked gas associated with outflows along the line of sight, in
  addition to the UV irradiated gas. The spectral resolution of the
  PACS and SPIRE spectrometers is insufficient to separate the PDR and
  shock contributions. The PDR models further suggest that optically
  thick ground-state water lines trace primarily surface layers of the
  PDR. The gas temperature in this region is predicted to be
  $\sim 800$~K and the water abundance with respect to hydrogen nuclei
  reaches $\sim 10^{-5}$. However, the bulk of gas-phase water
  is located in a more shielded region, deeper inside the PDR, with a
  gas temperature $\sim 24$~K, and a gas-phase H$_2$O abundance of
  $\sim 10^{-7}$. A comparison of the observed water line fluxes
  toward IRS1 with PDR model predictions shows no evidence for
  departures from the LTE ortho-para ratio at the local gas
  temperature, which is assumed in the models. The sensitivity of the
  \emph{Herschel} instruments was insufficient to measure the
  water ortho-para ratio toward the externally illuminated
  edge-on PDR at the surface of the S140 molecular cloud.
  Such observations will be possible with the next generation
  far-infrared spectroscopic space missions currently in development.}
  

\titlerunning{Water in PDRs}
\authorrunning{Lis et al.}

\maketitle 

\nolinenumbers

\section{Introduction}
\label{sec:intro}

\begin{figure*}
\sidecaption
\includegraphics[trim=0cm 5cm 0.1cm 0cm, clip=true, width=12cm]{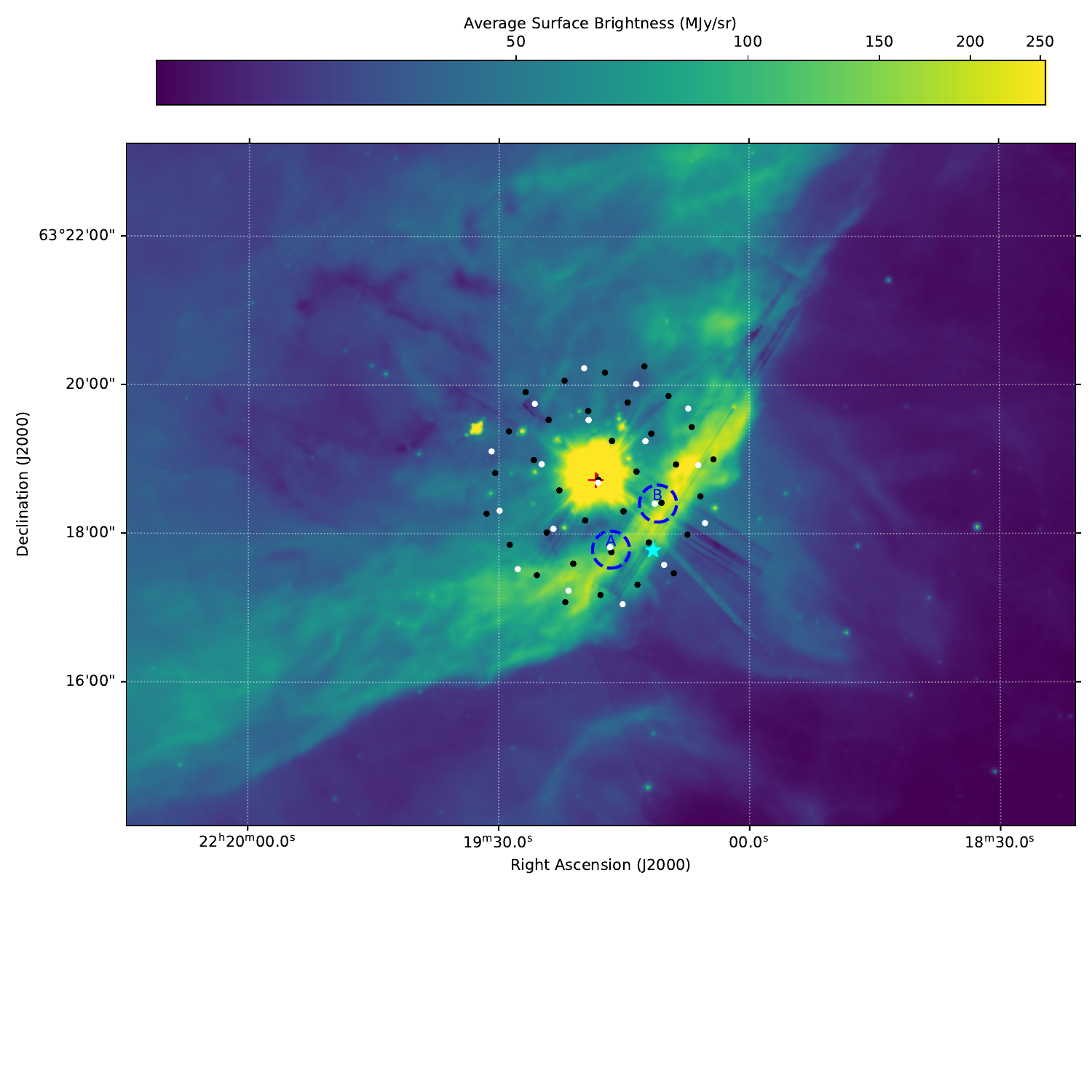}
\caption{\emph{Spitzer} IRAC Band 4 (8 \mic) image of the S140 region,
  dominated by PAH emission. The SPIRE SLW and SSW footprints are
  shown as white and black dots, respectively. The red cross marks the
  center of the PACS footprint toward the IRS1 protostar, the blue
  circles mark the positions A and B toward the surface PDR front, and
  the cyan star marks the \cite{Dedes2010}
  position. }\label{fig:Spitzer}
\end{figure*}

Understanding how the Earth obtained its water and whether water-rich
earth-like planets are common in the Universe is a fundamental
question in planetary formation research. Water forms efficiently in
the interstellar medium via gas-phase and grain-surface reactions and
is abundant in interstellar clouds as gas or ice in dust-grain
mantles. Gaseous water, along with water-ice covered dust grains,
either pristine or partially reprocessed, make their way through the
subsequent stages of star formation, including dense cores and
protostellar disks, before being incorporated into planetesimals in
forming planetary systems \citep{vanDishoeck2021}.

The possible links between the interstellar and Solar System materials
are of great interest for understanding the Solar System beginnings
\citep{Bockelee2000,Drozdovskaya2019}. The spin temperatures derived
from the ortho-para ratio (OPR) of water and other hydride molecules
with two spin isomers have often been interpreted as a measure of a
physical temperature at which the gas-phase molecules formed or
condensed on grain surfaces \citep{Crovisier1997,Kilaj2018}. This
makes the OPR a potentially important tracer of the thermal history of
water and of our interstellar heritage, in addition to molecular
abundances and isotopic ratios, such as the D/H and
$^{16}$O/$^{17}$O/$^{18}$O isotopic ratios in water, $^{14}$N/$^{15}$N
in ammonia, or $^{32}$S/$^{34}$S/$^{33}$S in H$_2$S
\citep{Sakamoto2007,Nomura2023,Goicoechea2021}.

The water spin temperatures measured in atmospheres of Solar System
comets are typically higher than $\sim 25$~K, corresponding to
OPR~$\ga 2$ \citep{Mumma2011}. The HIFI instrument on
\emph{Herschel} provided the first gas-phase measurements of the
water ortho-para ratio in diffuse and translucent interstellar medium
using absorption spectroscopy toward background dust continuum sources
\citep{Lis2010,Lis2013,Emprechtinger2013,Flagey2013}. The resulting
ortho-para ratios were generally found to be consistent with the
high-temperature limit of 3. OPR values below the high
temperature limit, but within the general range of cometary
measurements, have been reported toward Sagittarius~B2 in the Central
Molecular Zone ($2.34 \pm 0.25$ (2$\sigma$), spin temperatures
24--32~K; \citealt{Lis2010,Lis2013}), and toward some Galactic disk
clouds \citep{Flagey2013}. However, the overall measurement
uncertainties are large, as the resulting OPR values are sensitive to
the assumptions about the excitation of water molecules in the
foreground clouds. The standard assumption that all the ortho- and
para-water molecules are in the corresponding ground rotational states
has been demonstrated not to be satisfied in denser clouds, such as
those on the line of sight toward NGC!6334I \citep{Emprechtinger2013}.

Determination of the ortho-para ratio in regions, where water lines
are detected in emission, is much more sensitive to the assumptions
about the excitation and radiative transfer of water molecules
\citep{Lis2026}. Consequently, several very low OPR values initially
reported in some interstellar environments, corresponding to spin
temperatures of $10 - 20$~K, below the range of cometary measurements,
were subsequently revised. \cite{Hogerheijde2011} reported a low OPR
value of 0.77 in the TW~Hya disk. However, subsequent modeling
combining water and ammonia observations \citep{Salinas2016} showed
that the observations are consistent with a class of disk models
having water OPR values within the cometary range of 2.0--3.0.
Similarly, \cite{Choi2014} derived a water OPR of 0.1--0.5 in the
Orion Bar, unexpectedly low given the gas temperature in this region.
However, subsequent careful PDR modeling of the observations
\citep{Putaud2019} demonstrated that the low OPR reported by
\cite{Choi2014} was a result of a simplistic single temperature slab
model used in their analysis. The very low interstellar H$_2$O OPR
values reported in the literature have stimulated low-temperature
laboratory studies of the photodesorption of water ice. These
experiments showed that photodesorption releases H$_2$O into the gas
phase with an OPR of about 3 (e.g., \citealt{Hama2011, Hama2018}). In
contrast, it has been shown that the selection rules governing
gas-phase reactions involving the ortho and para forms of the
reactants can lead to low H$_2$O OPR values (\citealt{Faure2019},
  Michoud et al., in prep.)

Observations of water in PDRs are of particular interest, as
sophisticated models describing the physical and chemical structure of
these regions are available. Subsequent re-analysis of the water
observations in the Orion Bar PDR \citep{Putaud2019} clearly
demonstrated the need for a careful treatment of the molecular
excitation, radiative transfer, and chemical modeling when determining
accurate OPR values in regions, where water lines are seen in
emission, suggesting that the very low published OPR values may be
suspect.

As we plan observing programs with future FIR missions, such as the
NASA Astrophysics Probe, PRIMA, expected to launch in 2032, it is
critically important to improve and validate models needed to
accurately interpret OPR measurements in astrophysical environments,
going beyond simplistic single-slab LTE assumptions used in some
earlier analyses. This will ultimately determine how useful OPR
measurements are for understanding the water trail and the origins of
Solar System water. Namely, does it represent pristine
interstellar water, as suggested by the recent ALMA isotopic
measurements of the D/H ratio in a protoplanetary disk
\citep{Tobin2023}, or is it thermally and chemically reprocessed
during the formation of the Solar System and planets
\citep{Zannese2024}?

Here, we analyzed archival \emph{Herschel} spectroscopic
observations of water vapor and other molecular and atomic tracers in
S140, a well-studied PDR, to determine to what extent physical
conditions (density, temperature, and radiation environment) influence
the water formation pathways and abundances. We make use of the
InterStellar Medium DataBase (ISMDB) an online tool that enables easy
access to several thousands pre-computed chemical and physical models
for various ISM conditions and objects. The source is introduced in
Sect.~\ref{sec:s140}, the archival observations are presented in
Sect.~\ref{sec:obs}, the modeling results are described
Sect.~\ref{sec:discussion}, and the implications for the water
ortho-para ratio are discussed in Sect.~\ref{sec:water}.

\begin{figure}[ht!]
\centering
\includegraphics[trim=2cm 2.5cm 5.5cm 2cm, clip=true, 
   width=0.95\columnwidth]{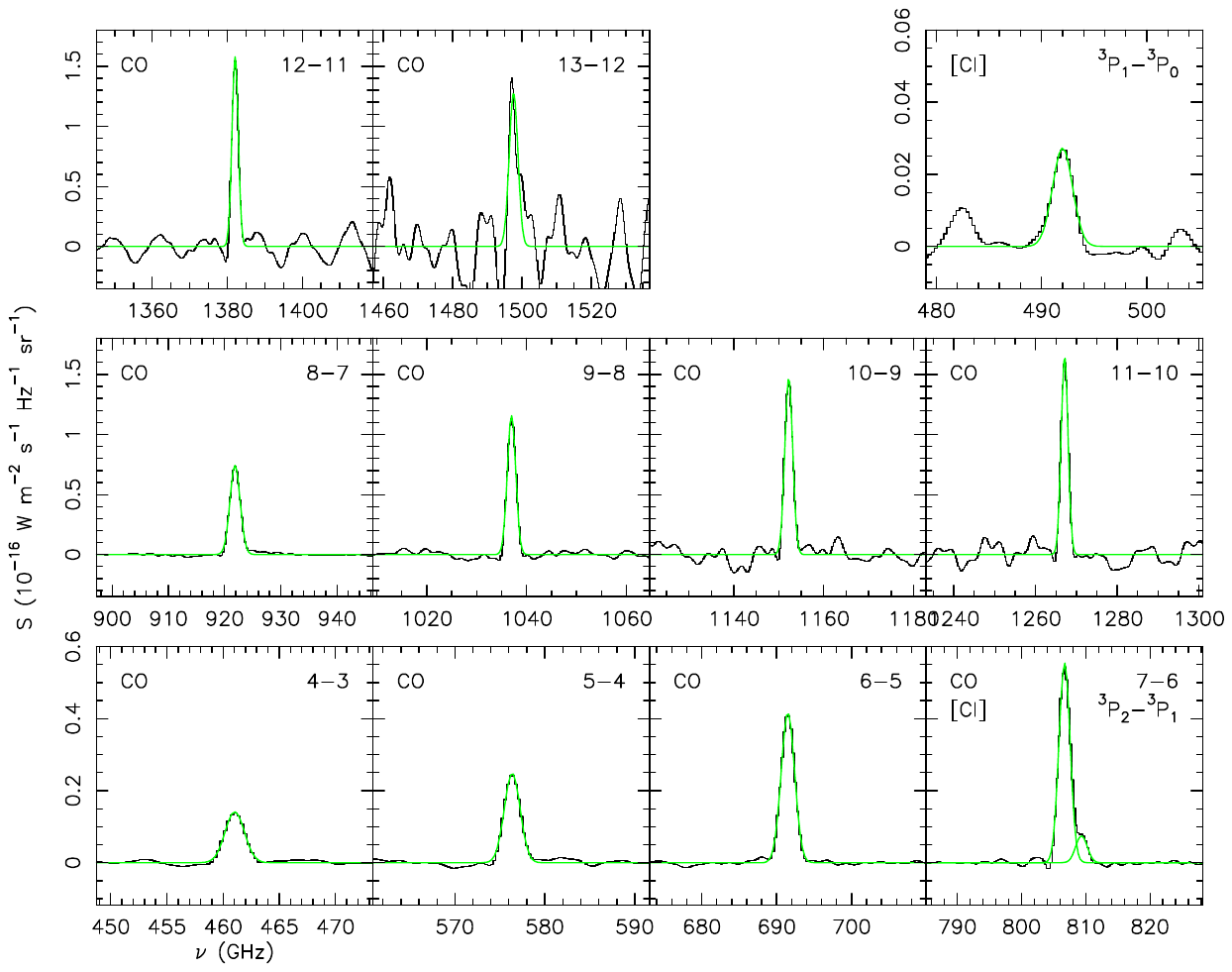} 
\includegraphics[trim=2cm 2.5cm 5.5cm 2cm, clip=true, 
   width=0.95\columnwidth]{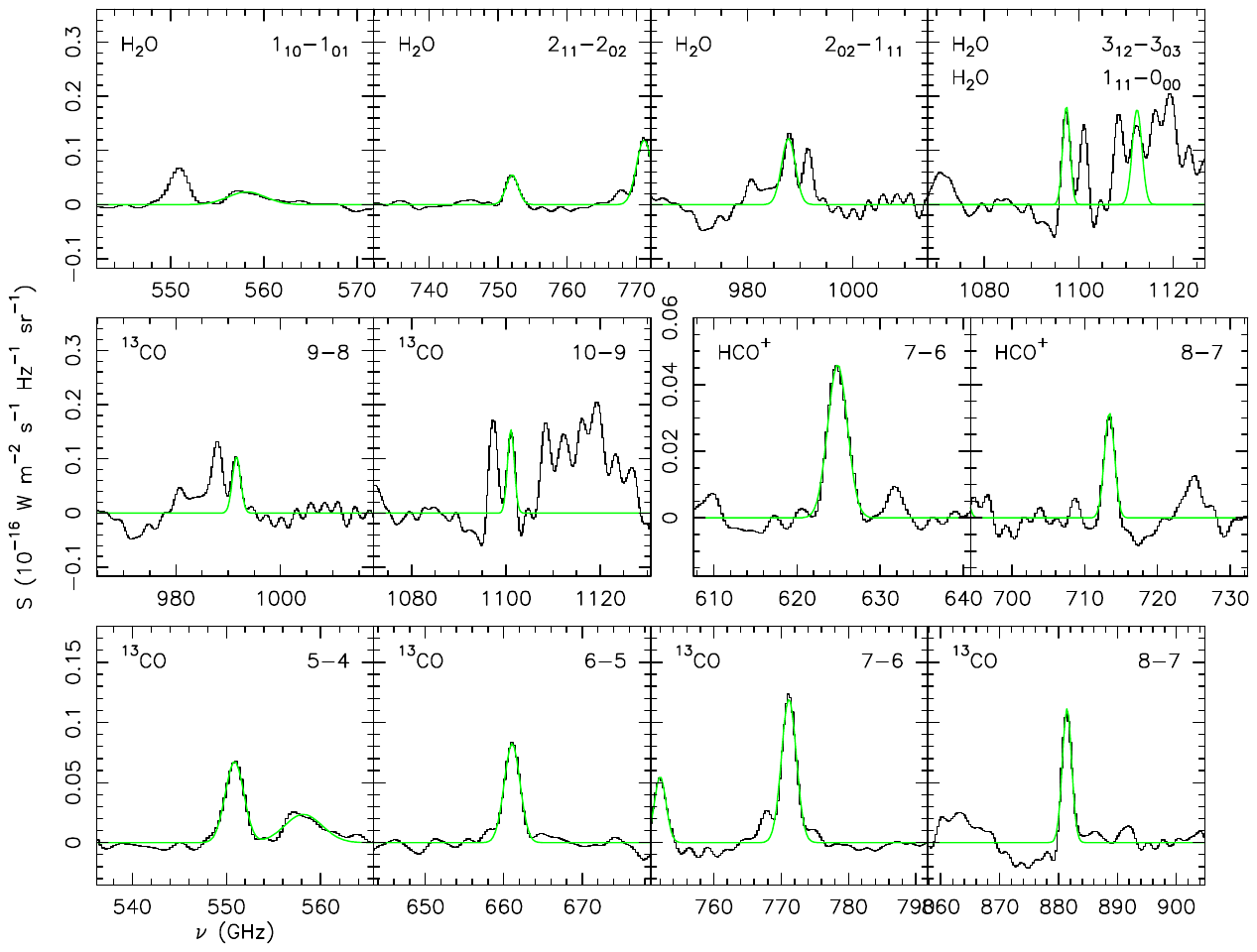} 
   \caption{SPIRE FTS spectra of CO, [C\,{\sc i}], H$_2$O, $^{13}$CO,
     and HCO$^+$ lines toward S140 IRS1.}\label{fig:spireco}
\end{figure}

\begin{figure}[ht!]
\centering
\includegraphics[trim=2cm 2.5cm 5.5cm 2cm, clip=true,
   width=0.95\columnwidth]{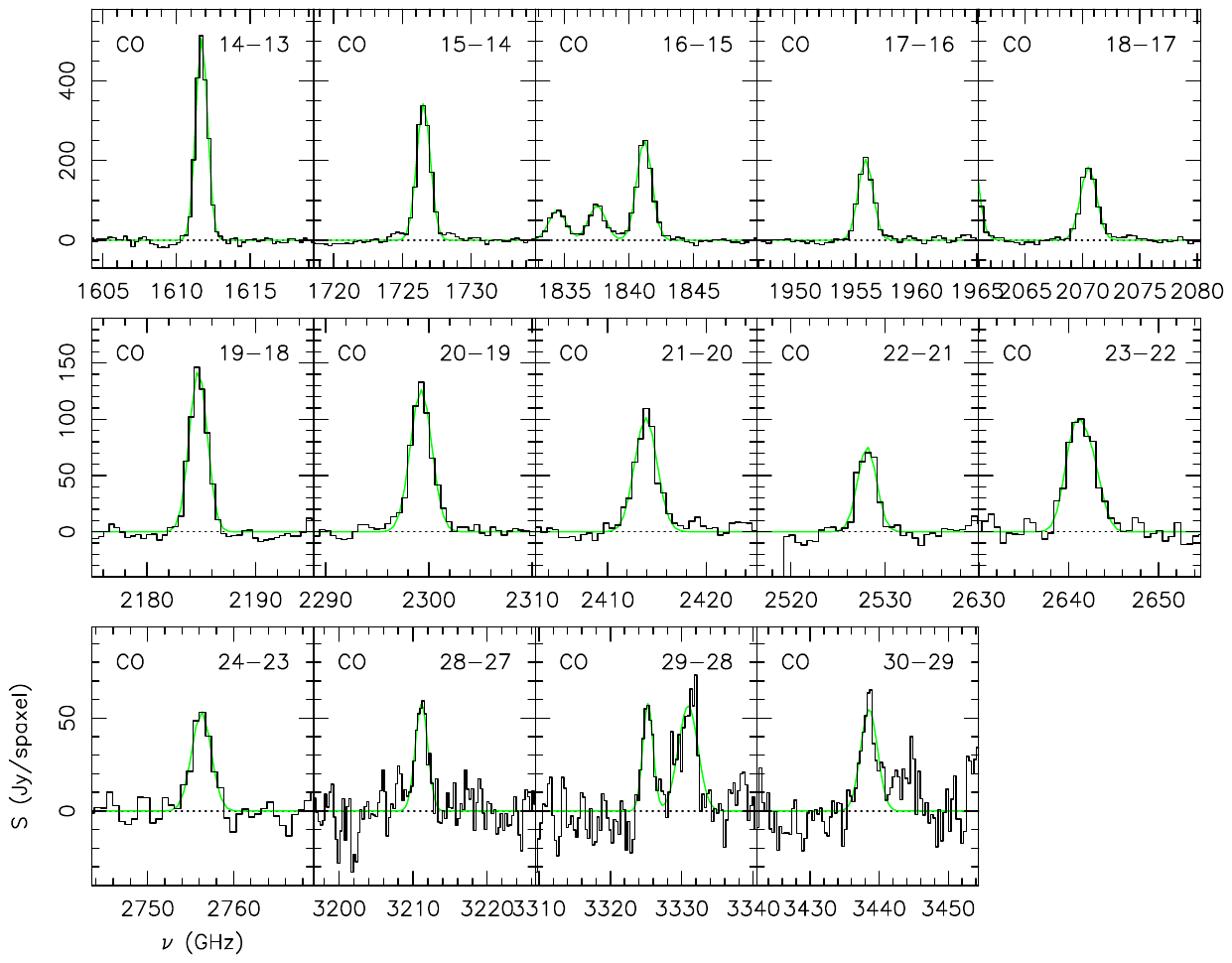} 
\includegraphics[trim=2cm 2.5cm 5.5cm 2cm, clip=true,
   width=0.95\columnwidth]{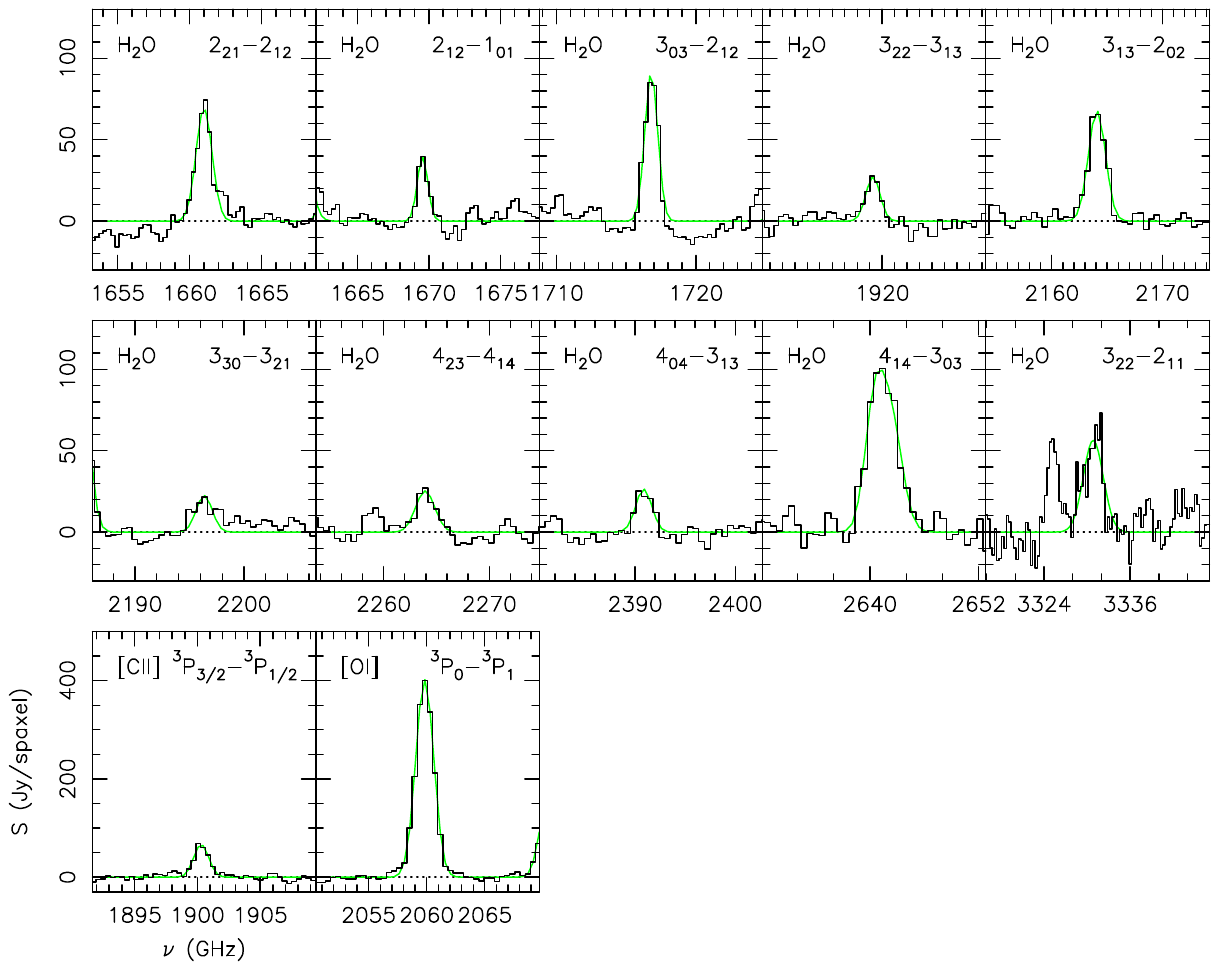} 
   \caption{PACS spectra of CO, H$_2$O, [C\,{\sc ii}], and [O\,{\sc
       i}] lines toward S140 IRS1.}\label{fig:pacsco}
\end{figure}

\begin{figure*}[ht!]
\sidecaption
\includegraphics[trim=2cm 3cm 5cm 1cm, clip=true, width=12cm]{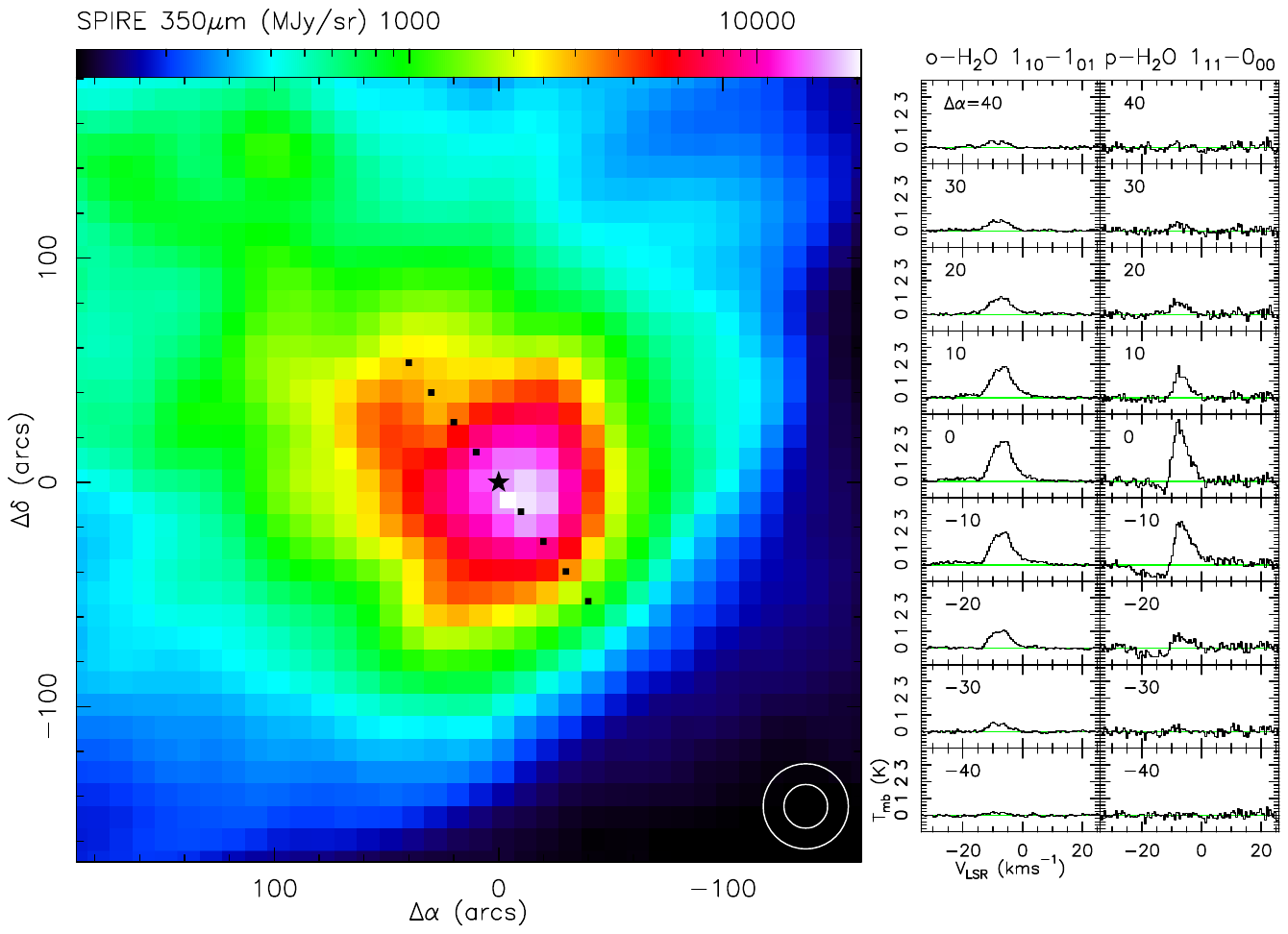}
\centering
\caption{HIFI spectra of the ground-state ortho- and para-water lines
  along a strip across the S140 PDR (right panel). The left panel
  shows the location of the strip superposed on the SPIRE 350~\mic\
  dust continuum image. The two white circles in the lower-right
  corner show the HIFI beam size at 557 and 1113~GHz, respectively.
  Offsets are in seconds of arc with respect to the HIFI
  pointing position.} \label{fig:hifi}
\end{figure*}

\section{S140 photon dominated region}
\label{sec:s140}

Located at 910~pc \citep{Crampton1974}, S140 is an HII region
\citep{Sharpless1959} and a molecular cloud bounded by a bright
H$\alpha$ rim \citep{Blair1978}. The surface of the cloud is
illuminated by the nearby B0.5V star HD\,211880 creating an almost
edge-on PDR. The core of the cloud contains a cluster of deeply
embedded objects \citep{Beichman1979}, including a massive protostar
S140~IRS1 \citep{Preibisch2002} driving a strong molecular outflow
\citep{Minchin1993}.

ISO observations of H$_2$ pure rotational and ro-vibrational
transitions as well as fine structure lines of [Fe\,{\sc ii}] and
[Si\,{\sc ii}] \citep{Timmermann1996} implied an average gas density
of $\sim 10^4$~cm$^{-3}$ and the incident UV flux of 400 Habing units
at the location of the surface PDR front.

Strip maps of the ground-state ortho-water and ammonia line emission
across the S140 region at 2\farcm1 spatial resolution were obtained by
\cite{Persson2009} using the \emph{Odin} satellite. The emission is
dominated by the outflow, however, it also shows a narrow line width
($\sim 3$~\kms) emission from an extended PDR. PDR modeling suggested
that the water emission arises mainly from surfaces of optically thick
clumps with H$_2$ densities in excess of $10^6$~cm$^{-3}$. The
resulting clump water abundance is $5 \times 10^{-8}$ with respect to
H$_2$, while the mean water abundance in the PDR is
$5 \times 10^{-9}$, and in the outflow between $\sim 4 \times 10^{-8}$
and $4 \times 10^{-7}$. The average ammonia abundance is about the
same as for water. Radiative transfer models of \cite{Poelman2006}
indicated that a combination of a pure PDR component and an embedded
source is required to match the low-resolution SWAS and ISO
observations of the water emission in S140.

\cite{Dedes2010} presented observations of [C\,{\sc ii}] and high-$J$
CO and HCO$^+$ emission in S140, at positions corresponding to the
externally irradiated surface PDR and the embedded massive protostar
IRS1. Their two-component clumpy PDR models toward IRS1 suggested the
UV illumination of $2.3 \times 10^5$ and 27 Draine units
($4 \times 10^5$ and 46 Habing units, respectively; 1 Draine unit
equals to 1.7 Habing units) for the hot and cool components,
respectively. The UV field of the cool component was lower than that
expected from the externally illuminating star HD~211880 (a few $10^3$
Draine units), suggesting that the edge-on PDR at the ionization front
blocks the UV radiation. Full PDR models were not feasible for the
surface PDR due to the small number of detected lines. However, the
mid- and high-$J$ CO and HCO$^+$ lines could only be explained by a
hot gas component with densities of order $10^6$~cm$^{-3}$ and a local
UV intensity of order $10^5$ Draine units, much higher than that
provided by the external radiation from HD~211880.

\section{Archival \emph{Herschel} observations}
\label{sec:obs}

\subsection{SPIRE observations}
\label{sec:obsspire}

S140 was observed using the SPIRE FTS instrument \citep{Griffin2010}
aboard the \emph{Herschel} Space Observatory
\citep{Pilbratt2010}\footnote{\emph{Herschel} was an ESA space
  observatory with science instruments provided by European-led
  Principal Investigator Consortia and with important participation
  from NASA.} as a part of the guaranteed time program {\tt
  KPGT\_epoleham\_1} (\emph{Herschel} OBSID 1342212319). Observations
were centered toward the IRS1 position
$\alpha_{\rm J2000}=22^{\rm h} 19^{\rm m} 18\fs07$,
$\delta_{\rm J2000} = 63\arcdeg\ 18\arcm\ 43\farcs65$. Additional
information is provided in Appendix~\ref{app:obs}.

Figure~\ref{fig:spireco} shows spectral lines detected with SPIRE FTS
toward S140 IRS1. This includes ten CO lines, six $^{13}$CO lines,
four water lines, two [C\,{\sc i}] fine structure lines (the 809~GHz
line is partially blended with CO), and two HCO$^+$ lines. The
corresponding line fluxes are listed in Table~\ref{tab:spire}.

\subsection{PACS observations}
\label{sec:obspacs}

S140 was observed using the PACS instrument \citep{Poglitsch2010}
aboard \emph{Herschel} as a part of the guaranteed time program {\tt
  KPGT\_vossenko\_1} (\emph{Herschel} OBSID 1342222256). Observations
were taken using the PACS Range Spectroscopy observing mode, centered
toward the IRS1 position
$\alpha_{\rm J2000}=22^{\rm h} 19^{\rm m} 18\fs24$,
$\delta_{\rm J2000} = 63\arcdeg\ 18\arcm\ 44\farcs92$ (red cross in
Figure~\ref{fig:Spitzer}). Additional information is provided in
Appendix~\ref{app:obs}. 

PACS CO and H$_2$O
spectra toward S140 IRS1 are shown in Figure~\ref{fig:pacsco}.
Fourteen CO transitions between $J= 14-13$ and $30-29$ and ten water
lines are detected and used in the subsequent analysis. The fine
structure lines of [C\,{\sc ii}] and [O\,{\sc i}] are also detected.

\begin{figure}[ht!]
\centering
\includegraphics[trim=1.3cm 2.5cm 6cm 3cm, clip=true, width=0.95\columnwidth]{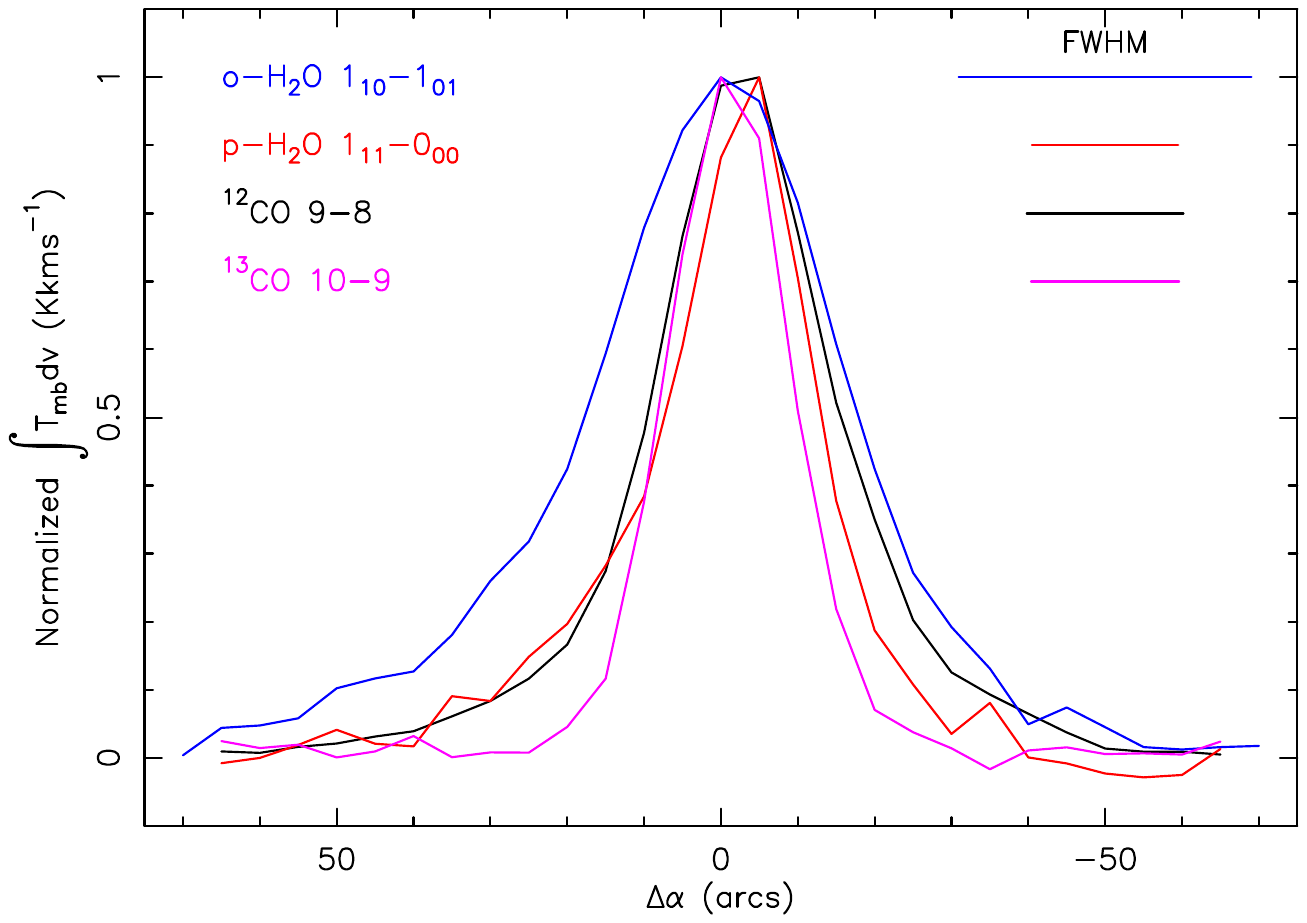} 
\caption{Normalized integrated line intensities of CO, $^{13}$CO, and
  water lines along a strip across the S140 PDR. The corresponding
  HIFI beam sizes are shown in the upper-right corner. Given the
  relatively large distance to the source, only a small spatial
  stratification is seen.} \label{fig:strip}
\end{figure}

\subsection{HIFI observations}
\label{sec:obshifi}

S140 was observed using the HIFI instrument \citep{deGraauw2010}
aboard \emph{Herschel} as a part of the guaranteed time program {\tt
  KPGT\_vossenko\_1} (\emph{Herschel} OBSIDs 1342201741, 1342201806,
1342244407). On-the-fly strips with a length of $\sim 215\arcs$
perpendicular to the ionization front were observed at the frequencies
of the CO~9--8, $^{13}$CO 10--9, as well as $p-$H$_2$O $1_{11}-0_{00}$
and $o-$H$_2$O $1_{10}-1_{01}$ lines. The strip was centered at
$\alpha_{\rm J2000}=22^{\rm h} 19^{\rm m} 18\fs88$,
$\delta_{\rm J2000} = 63\arcdeg\ 18\arcm\ 52\farcs9$.

Figure~\ref{fig:hifi} shows the position of the strip superposed on
the SPIRE 350~\mic\ dust continuum emission, as well as the resulting
spectra of the ground-state ortho- and para-water lines.
Figure~\ref{fig:strip} shows the integrated line intensity of the
molecular tracers along the strip, normalized to their corresponding
peaks as a function of the right ascension offset. At the angular
resolution of \emph{Herschel}, the spatial stratification is not as
clearly visible in S140 as in nearby edge-on PDRs. The
ground-state ortho-water line peaks further away from the ionizing
star than $^{12}$CO, and close to the $^{13}$CO peak, while the
ground-state para-water line coincides with the $^{12}$CO peak.
However, these shifts are likely dominated by the complex morphology
of the IRS1 region, including local heating from the embedded
protostar, and variations in the beam size, which differs by almost a
factor of 2 between the two water transitions. In plane parallel PDR
models (Sec.~\ref{sec:discussion}) the local emissivity of the ground
state ortho- and para-water lines peaks over approximately the same
range of $A_V$ values, $\sim 0.25 - 0.5$ in the optically thin case,
shifting to a narrow range around $A_V \simeq 0.25$ when the line
optical depth is taken into account.

Velocity-resolved HIFI CO~9--8 and $o-$H$_2$O $1_{10}-1_{01}$ spectra
toward IRS1 show evidence of high-velocity emission associated with
outflows. The exact line shapes depend on the optical depth and
excitation requirements, which are different for different
transitions. However, high-velocity wings are clearly visible. A
simple two-component Gaussian fit gives $\sim 13$ and 27~km\,s$^{-1}$
line-widths of the broad component for CO and H$_2$O, respectively. In
the case of the $o-$H$_2$O $1_{10}-1_{01}$ water line, $\sim 30$\% of
the emission comes from the broad component. Although \emph{Herschel}
observations show that the water abundance in outflows and shocks is
moderately low (\citealt{vanDishoeck2021} and references
therein), such regions likely contribute to the observed water
line fluxes toward IRS1. Given the low spectral resolution of the PACS
and SPIRE spectrometers, separating the narrow and broad-component
emission is not possible for other lines. Consequently, we use
integrated line fluxes in the analysis.

Given the much stronger dust continuum emission at 1113~GHz compared
to 557~GHz, the high-velocity gas at negative velocities extending
down to $\sim -25$~km\,s$^{-1}$ is seen in absorption in the
$p-$H$_2$O $1_{11}-0_{00}$ line (Fig~\ref{fig:hifi}). The blue lobe
of the outflow is thus located in front of IRS1.

\section{PDR modeling}
\label{sec:discussion}

The Meudon PDR code \citep{LePetit2006} is a publicly available,
state-of-the-art astrochemistry code that computes self-consistently
the radiative transfer, the chemistry both in the gas phase and on
grain surfaces, as well as the thermal processes and thermal balance
at each point within a UV-irradiated cloud. The code simulates a
stationary plane-parallel slab of gas and dust illuminated on both
sides by the radiation field from the far UV (FUV) to the radio
domain. For the illuminating UV radiation field, the code can 
consider the ISRF description from  \cite{Mathis1983}, \cite{Draine1978},
or specific stellar spectra.  The FUV part of the ISRF that  affects 
photodissociation is the Lyman-Werner band, 912--1108~\AA. The flux 
of those photons is controlled by the O and B star populations.

For the chemistry, the code considers approximately 200 chemical
species linked by a chemical network of $\sim 2000$ chemical
reactions. For the grain surface chemistry, a power-law grain size
distribution is considered. The surface processes that are considered
are the adsorption from the gas, reactions in the bulk and on the
surface considering thermal and tunneling effects, desorption
processes (thermal and non-thermal photo-desorption, cosmic ray
desorption, chemical desorption, i.e., release of molecules into the
gas phase due to the exothermicity of some reactions).

A particularly useful online tool is the InterStellar Medium DataBase
(ISMDB)\footnote{http://ismdb.obspm.fr. The latest version of the
  ISMDB database (grid ID P71PDRE20v05\_P\_250918 and
  P71PDRE20v05\_n\_250918) is used in the models presented here.},
which allows quick and easy access to pre-computed PDR and
shock models. The grid of models computed using the Meudon PDR code is
used in the present study. Presently, ISMDB gives access to
several thousand Meudon PDR models covering a wide range of
astrophysical conditions. To prepare and interpret observations, ISMDB
provides an inverse search service to allow queries on observables via
a Jupyter notebook, to find the best model parameters that match
observations and gives access to line intensity maps in a parameter
space (e.g., density, UV flux). The best-fitting model is
determined by minimizing the $\chi^2$. ISMDB supports both additive and
multiplicative uncertainties. Depending on which type of uncertainty
dominates, the $\chi^2$ is computed either in linear space (for
additive uncertainties) or in logarithmic space (for multiplicative
uncertainties). Once the best fit model is found, model results
can be downloaded for further analysis.

\begin{figure}
\centering
\includegraphics[trim=0cm 0cm 0cm 0cm, clip=true, width=0.95\columnwidth]{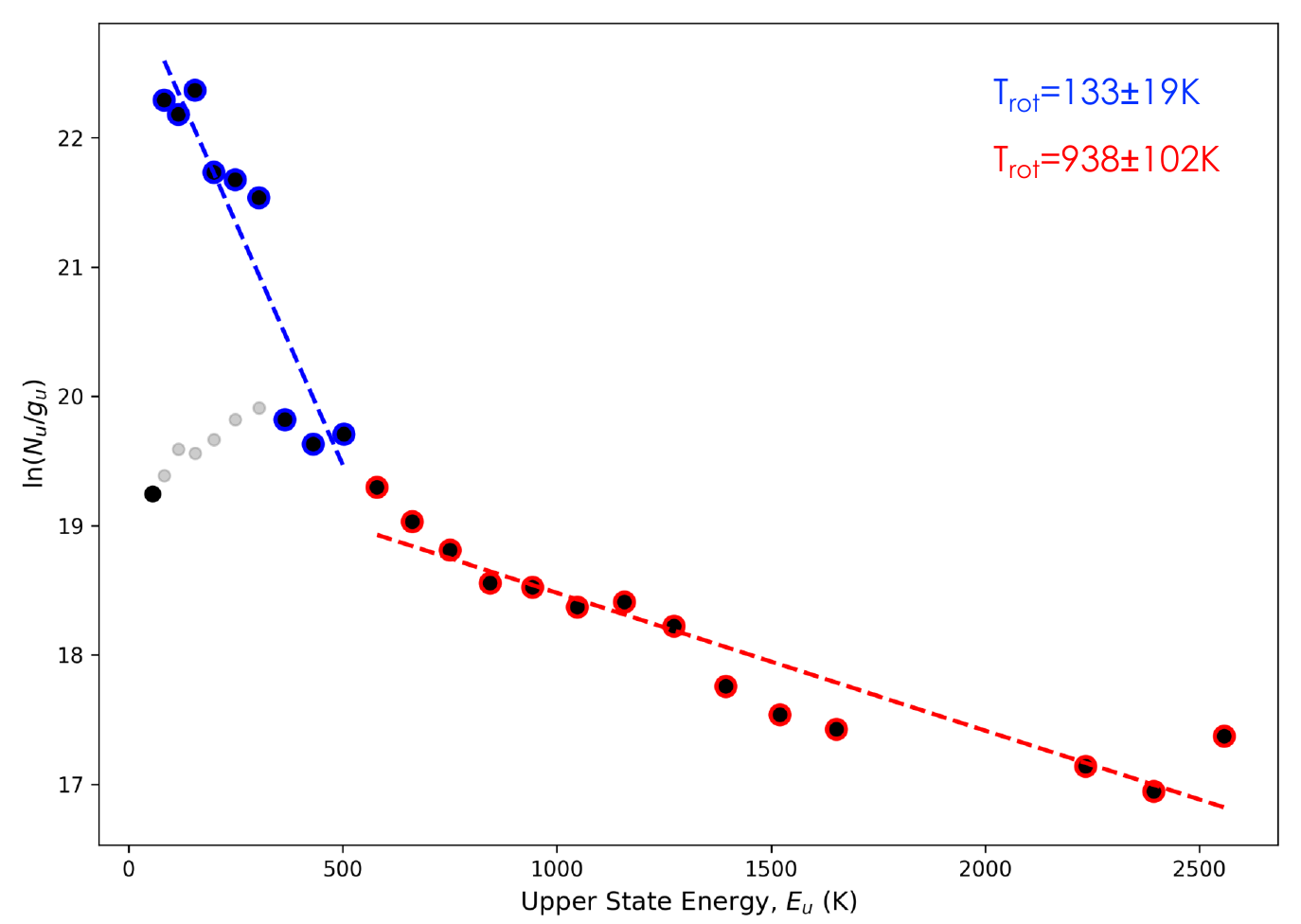} 
\caption{CO rotation diagram toward IRS1. Gray and blue points
  correspond to uncorrected and opacity-corrected line fluxes. The
  dashed lines show least square fits to the points with $E_u$ below
  and above 500~K. The opacity correction is not possible for the
  lowest CO $J=4-3$ transition, as no $^{13}$CO data are available.
  This transition is thus excluded from the fit.}\label{fig:rotdiag}
\end{figure}

\begin{figure*}[!ht]
\sidecaption
\includegraphics[trim=0cm 0cm 0cm 0cm, clip=true, width=12cm]{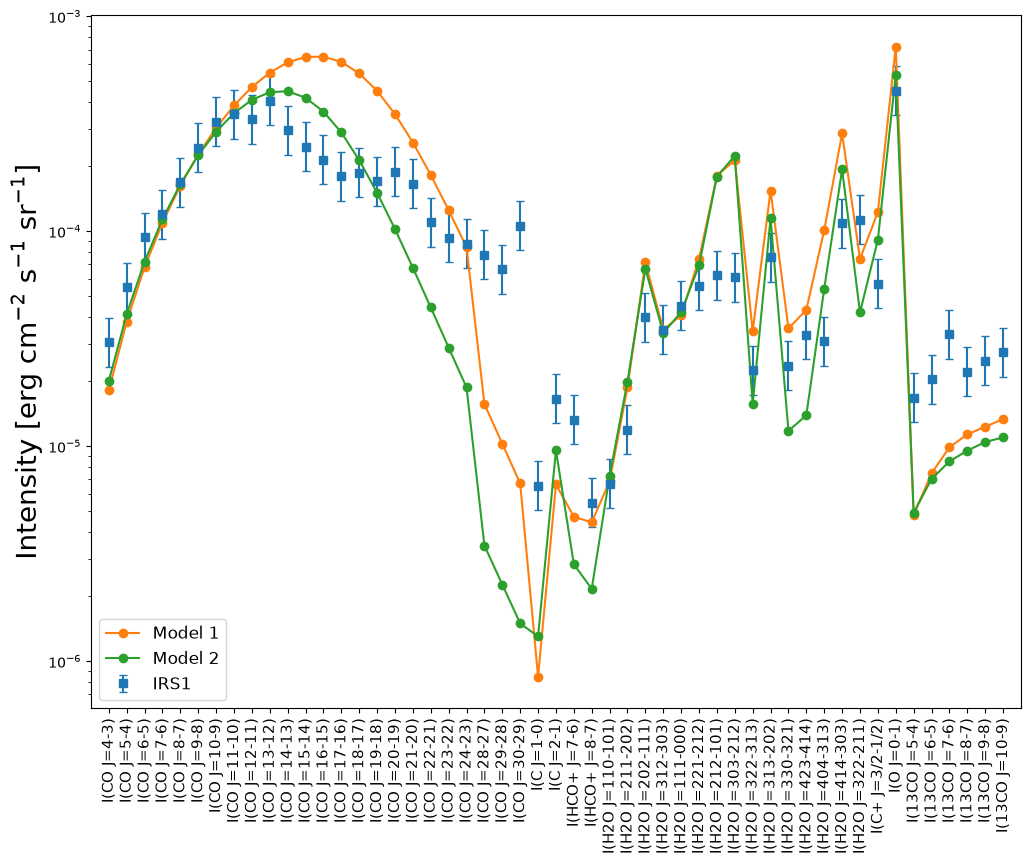}
\caption{Best fit isobaric ISMDB model of all SPIRE and PACS
  observations toward S140~IRS1 (Model 1, orange), and the preferred
  model excluding high-$J$ CO and $^{13}$CO observations, as described
  in the text (Model 2, green).}\label{fig:ismdb}
\end{figure*}

\subsection{CO rotation diagram toward S140 IRS1}
\label{sec:rotdiag}

The HIFI spectra toward IRS1 show broad line widths
($13 - 27$~km\,s$^{-1}$ ), indicating the presence of outflows
(Fig.~\ref{fig:hifi}). Consequently, shocks may be present along the
line of sight, affecting the shape of the CO SLED. We employed the
rotation diagram approach to compare the CO observations of IRS1 with
similar regions from the literature.

We used the rotation diagram approach for CO following
\cite{Goicoechea2015} to estimate the rotational temperature
($T_{rot}$) of the molecular gas. For each CO transition, the
upper-level column density was calculated using the formula:

\begin{equation}
  N_u = {\frac{8 \pi k \nu^2 W}{h c^3 A_{ul}}}
\end{equation}
 
\noindent where $\nu$ is the transition frequency, $W$ is the
integrated intensity, $A_{ul}$ is the Einstein spontaneous emission
coefficient, $k$ is the Boltzmann constant, $h$ is Planck constant,
and $c$ is the speed of light, all in CGS units. The quantity
$\ln ( N_u / g_u)$, where $g_u$ was the statistical weight of the
upper level, is then plotted against the upper-level energy $E_u$
(Figure~\ref{fig:rotdiag}). The slope of the linear fit to the data
corresponds to $-1/T_{rot}$.

We applied a correction for the optical depth of the low-$J$ $^{12}$CO
lines by comparing $^{12}$CO and $^{13}$CO line intensities assuming a
$^{12}$C/$^{13}$C isotopic ratio of 60. We multiplied the upper-level
column density $N_u$ by the corresponding correction factors. This
technique assumes local thermodynamic equilibrium (LTE).

The CO rotation diagram toward S140 IRS1 (Figure~\ref{fig:rotdiag}) is
well described by two components with rotational temperatures of
$133 \pm 19$ and $938 \pm 102$~K, respectively. \cite{Goicoechea2015}
derived similar rotational temperatures of $T_{rot} \sim 370$~K and
$\sim 1000$~K toward the Orion Peak 1 shock, with the ambient PDR gas
being characterized by a rotational temperature of $\sim 180$~K.
\cite{vanDishoeck2021} summarized \emph{Herschel} observations of the
CO SLEDs from low- to high-mass protostars. Their Figure~14 shows that
two-component rotation diagrams with excitation temperatures of
$T_{rot} \sim 300$ and $\sim 700$~K are typical across the protostar
mass spectrum. In fact, for many protostars, three temperature ranges
can be identified in CO excitation diagrams: (i) low-$J$ (Jup < 14)
probing warm quiescent and entrained outflow gas; (ii) mid-$J$
($J_{up} = 14 - 24$) probing the 300 K broad component; and (iii)
high-$J$ ($J_{up} > 24$) probing the hot 700 K shock component.

Our CO rotational diagram toward S140~IRS1, reveals two gas components
with rotational temperatures of $\sim 130$ and $\sim 940$~K, which we
associate with the PDR and shocked gas, respectively. The S140~IRS1 CO
SLED is markedly different from that observed toward the Orion Bar,
which shows a single component well described by a rotational
temperature of $147 \pm 9$~K (see Fig.~4 of \citealt{Joblin2018}).
Detailed shock modeling aimed at better reproducing the CO SLED is
beyond the scope of the present paper.

\subsection{ISMDB models of S140 IRS1}

We used the SPIRE and PACS observations toward S140 IRS1 (listed in
Tables~\ref{tab:spire} and \ref{tab:pacs}) to characterize the
physical conditions in this source from the ISMDB fit. The [N\,{\sc
  ii}] 205~$\mu$m fine structure line, which is a tracer of
photoionized gas in H\,{\sc ii} regions is excluded, as it is not
modeled by the Meudon code. A 30\% calibration uncertainty was added
in quadrature to the statistical uncertainty computed for each line
from the rms in the surrounding emission-free channels. We considered
isobaric (constant pressure) models that have been shown in earlier
studies to better reproduce PDR observations (e.g.,
\citealt{Putaud2019}). The basic model parameters are: thermal
pressure $p_{th}$, interstellar radiation field enhancement factor
$G_0$, and thickness of the slab expressed as visual extinction $A_V$.

To account for geometric effects related to coupling between the
source and the telescope beam when comparing our plane-parallel models
with observations of a structured PDR surface, we introduced as an
additional free parameter a geometrical scaling factor, $\Omega$,
which has the same value for all lines used in the fit. This factor
encapsulates the combined influence of several effects, namely the
possible presence of multiple PDR layers along the line of sight, the
inclination of the emitting surface, the beam-filling factor, and the
angle of incidence of the UV radiation. It is used to scale the
face-on line intensities predicted by the PDR models, following the
approach of \cite{Sheffer2011}. The Meudon PDR code allows computing
emergent line fluxes for different viewing inclination angles between
the line of sight to the illuminating star and the observer. Model
line fluxes used in the analysis were computed assuming a viewing
inclination angle of zero (i.e., normal viewing angle).

As a starting point of the analysis, we searched for a pure PDR
model best reproducing the intensities of all the SPIRE and PACS lines
observed in S140 IRS~1. The result is represented in
Figure~\ref{fig:ismdb} as orange points (Model 1). The model has a
pressure $6.7 \times 10^8$~K\,cm$^{-3}$, a radiation field scaling
factor $G_0=3.5 \times 10^3$, a visual extinction $A_V=40$, and a
geometrical scaling factor $\Omega=0.25$. The observed line fluxes
typically to within a factor of $\sim 3$. This agreement can be
considered satisfactory for a complex source like S140~IRS1, given the
simplifying model assumptions of plane-parallel geometry and normal UV
incidence angle. However, there are some notable discrepancies between
the model and the observations. The model fails to accurately
reproduce the shape of the CO spectral line energy distribution
(SLED), over-predicting fluxes of the mid-$J$ CO lines, between
$J=14-13$ and $23-22$ and under-predicting fluxes of lines above
$J=28-27$. The difference in the shape of the CO SLED between the
model and observation further suggests that processes other than UV
illumination are at play in this region (see Sect.~\ref{sec:rotdiag}).
This interpretation is consistent with the high excitation
temperature determined from the rotation diagram, which could indicate
the presence of shocks. The model also under-predicts the intensities
of the $^{13}$CO and [C\,{\sc i}] lines. As suggested by
\cite{Joblin2018}, this may be due to the self-shielding of the
$^{13}$CO lines by H$_2$ and $^{12}$CO lines not being included in the
grid of models published in ISMDB.

To further consider the hypothesis that shocks contribute to the
highly excited CO emission, we searched for an alternative model that
fits the observed lines, with a specific criterion for CO: the model
was required to reproduce the intensities of CO lines up to $J=13-12$ while
under-predicting those from $J=15-14$ and above, which were
assumed to contain an additional contribution from shocks.
The $^{13}$CO lines were also excluded from the fit for the reasons
discussed above. The resulting model is shown as green points (Model
2) in Fig.~\ref{fig:ismdb}. It has a pressure
$5 \times 10^8$~K\,cm$^{-3}$, a lower radiation field scaling factor
$G_0=1 \times 10^3$, a visual extinction $A_V = 20$, and a geometrical
scaling factor $\Omega = 0.75$. As expected, high-$J$ CO lines are
strongly under-predicted, but differences for other lines are modest.
We use this model as our preferred model in the subsequent analysis.

\cite{Dedes2010} presented a two-component clumpy PDR model for
S140~IRS1. They derived densities of the hot and cold components of
$1.8 \times 10^6$~cm$^{-3}$ and $1.3 \times 10^6$~cm$^{-3}$,
respectively, with the corresponding $G_0$ values of $4 \times 10^5$
and 46 Habing units, respectively. Thus, the $G_0$ derived here is
intermediate between the values for their hot and cold components. We
note that isobaric models considered here create naturally a density
gradient across the PDR and, as suggested by \cite{Marconi1988} and
\cite{Allers2005}, provide a better physical description compared to
the clumpy PDR approach.

The pressure for our preferred isobaric model of S140 IRS1 is
similar to the value derived for the high-UV illumination PDR, the
Orion Bar, $5 \times 10^8$~K\,cm$^{-3}$ vs.
$3.5 \times 10^8$~K\,cm$^{-3}$, while $G_0$ is an order of magnitude
lower $1 \times 10^3$ vs. $1.9 \times 10^4$, respectively
\citep{Putaud2019}. This is consistent with IRS1 being a high-mass
protostar with a luminosity of $8.5 \times 10^3$~\lsun\
\citep{Maud2013}. The model parameters should be considered as
indicative rather than uniquely constrained, owing to the degeneracy
between the pressure and $G_0$ values. Models with somewhat lower
pressures and higher $G_0$ can provide acceptable fits to the
observations. A more robust determination of the physical conditions
would require a self-consistent treatment combining PDR and shock
components, so that their respective contributions to the observed
line emission can be constrained simultaneously.

\subsection{ISMDB models of the surface PDR}
\label{sec:ifront}

The S140 surface PDR front is covered by the the off-center pixels of
the SLW and SSW arrays. We selected for the analysis two positions
along the ionization front for which SLW and SSW pixels coincide.
These positions, labeled A and B (blue circles in
Figure~\ref{fig:Spitzer}), are located on both sides of the
\cite{Dedes2010} position. The line fluxes of the detected transitions
are listed in Table~\ref{tab:spireif}, along with the average of the
two positions. We also included the [C\,{\sc ii}] line flux from
\cite{Dedes2010} along with the average line fluxes at positions A and
B, to better constrain the models.

\begin{figure}[ht!]
\centering
\includegraphics[trim=0cm 0cm 0cm 0cm, clip=true, 
  width=0.95\columnwidth]{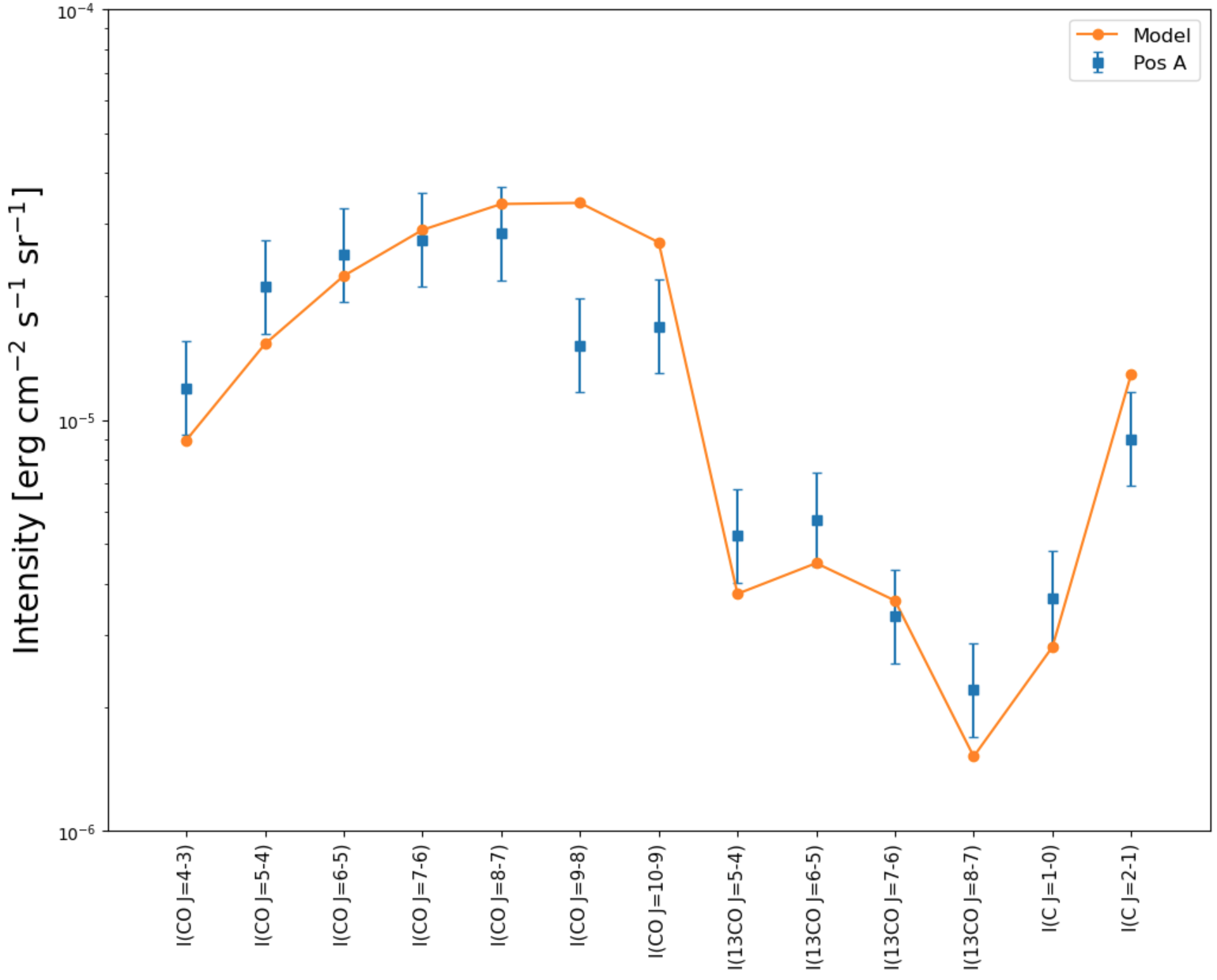} 
\includegraphics[trim=0cm 0cm 0cm 0cm, clip=true, 
  width=0.95\columnwidth]{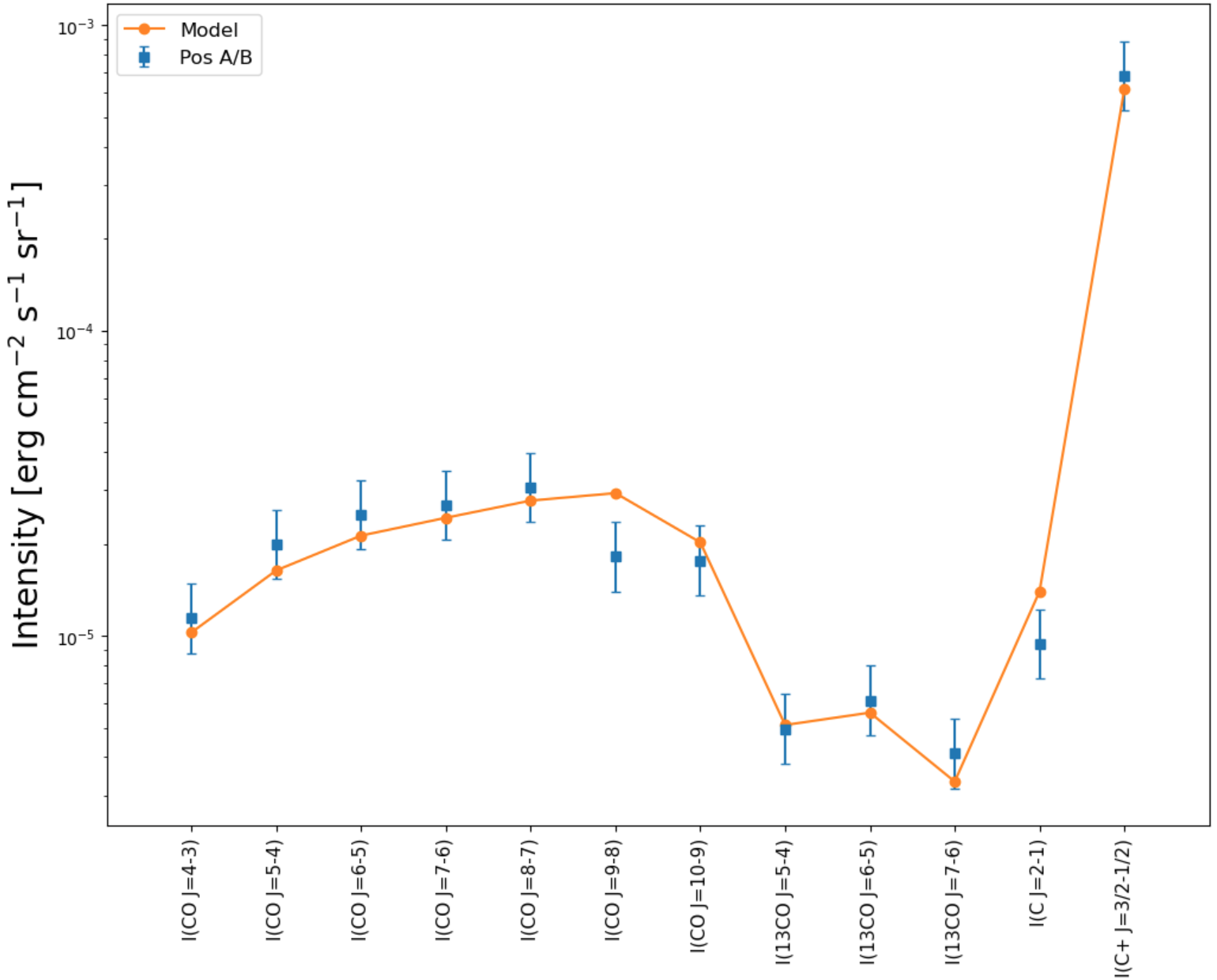} 
  \caption{Best fit isobaric ISMDB models of the SPIRE observations
    toward the S140 surface PDR, positions A (top panel) and average
    of positions A and B, including the [C\,{\sc ii}] measurement of
    \cite{Dedes2010} (lower panel).}\label{fig:ismdbif}
\end{figure}

Our best-fit isobaric models at the position A and the average of A/B
(Fig.~\ref{fig:ismdbif}) yield a pressure of $1 \times 10^6$
K\,cm$^{-3}$ and $G_0$ of $1.9\times 10^4$. \cite{Dedes2010} attribute
the emission in the high-$J$ lines of CO and HCO$^+$ to a hot gas
component with densities on the order of $1.0 \times 10^6$ cm$^{-3}$
and $G_0$ on the order of $10^5$. However, much lower $G_0$
values at the surface PDR have also been reported in the literature.
\cite{Ossenkopf2015} derived a very weak UV field of 60 Draine units
based on observations of the [O\,{\sc i}] 63 $\mu$m and [C\,{\sc ii}]
158~$\mu$m fine structure lines, and CO 13--12 and 16--15 rotational
lines. \cite{Timmermann1996} estimated the UV strength at the PDR
surface to be 400 Habing units based on observations of H$_2$ pure
rotational and ro-vibrational transitions, as well as [Fe\,{\sc ii}],
and [Si\,{\sc ii}] fine structure lines. The resulting $G_0$ thus
strongly depend on the data set and the modeling approach used in the
analysis. Our best-fit G$_0$ value, based on a different data set and
different analysis, is in the middle of the broad range of values
reported in the literature. The wide range of $G_0$ values reported in
the literature highlights the need for a dedicated study combining all
the available observational constraints using both PDR and shock
models.

The presence of CO at the S140 surface PDR can be explained by an
isobaric PDR, as observed in other regions such as the Orion Bar, NGC
7023 \citep{Joblin2018}, and Carina \citep{Wu2018}. The underlying
physical processes are as follows. The gas in this region is
compressed to a high thermal pressure through photoevaporation driven
by UV photons \citep{Bron2018}. In the resulting dense, warm, and
strongly UV-irradiated layer, H$_2$ molecules are electronically
excited by the UV pumping and subsequently decay to the ground-state
via fluorescence, followed by cascades through ro-vibrational levels.
The internal ro-vibrational energy of H$_2$, together with the
elevated gas temperature, enables reactions that overcome activation
barriers, leading to the formation of species such as CH$^+$ and OH
\citep{Goicoechea2017}. These then participate in chemical networks
that ultimately produce HCO$^+$ and CO \citep{Goicoechea2016}.

\section{Water ortho-para ratio and abundance in S140 IRS1}
\label{sec:water}

No water lines were detected in the SPIRE FTS spectra toward the
externally illuminated edge-on PDR on the surface of the S140
molecular cloud, where the CO lines are typically an order of
magnitude weaker than those observed toward IRS1. However, eight
ortho- and seven para-water lines were detected toward IRS1, which
allows us to study the water ortho-para ratio on this line of sight.
Table~\ref{tab:water} lists the observed line fluxes of the ortho- and
para-water lines detected toward S140~IRS1 compared to those predicted
by the best fit isobaric model to all the PACS and SPIRE observations
described above.

\begin{table*}
\begin{center}  
\caption{Observed and modeled water line fluxes toward S140 IRS1.} 
\label{tab:water}
\begin{tabular}{lccccc}
\hline \hline 
  \rule[-3mm]{0mm}{8mm}  Line & $\nu$ & $E_l$ & $S_{o}$ & $S_{m}$  & $\log (S_{o}/S_{m})$\\

  \hline 
$o-$H$_2$O	$1_{10}-1_{01}$	&	556.936	        &	34.2	&	6.68(-6)	&	6.94(-6)	&	-0.02	\\
$o-$H$_2$O	$3_{12}-3_{03}$	&	1097.3648	&	196.8	&	3.48(-5)	&	3.21(-5)	&	0.04	\\
$o-$H$_2$O	$2_{21}-2_{12}$	&	1661.0076	&	114.4	&	5.58(-5)	&	6.61(-5)	&	-0.07	\\
$o-$H$_2$O	$2_{12}-1_{01}$	&	1669.9048	&	34.2	&	6.25(-5)	&	1.72(-4)	&	-0.44	\\
$o-$H$_2$O	$3_{03}-2_{12}$	&	1716.7696	&	114.4	&	6.11(-5)	&	2.14(-4)	&	-0.54	\\
$o-$H$_2$O	$3_{30}-3_{21}$	&	2196.3458	&	305.2	&	2.37(-5)	&	1.12(-5)	&	0.32	\\
$o-$H$_2$O	$4_{23}-4_{14}$	&	2264.1497	&	323.5	&	3.30(-5)	&	1.33(-5)	&	0.39	\\
$o-$H$_2$O	$4_{14}-3_{03}$	&	2640.4738	&	196.8	&	1.09(-4)	&	1.85(-4)	&	-0.23	\\
  \hline												
$p-$H$_2$O	$2_{11}-2_{02}$	&	752.0331	&	100.9	&	1.19(-5)	&	1.91(-5)	&	-0.21	\\
$p-$H$_2$O	$2_{02}-1_{11}$	&	987.9268	&	53.4	&	3.98(-5)	&	6.39(-5)	&	-0.21	\\
$p-$H$_2$O	$1_{11}-0_{00}$	&	1113.343	&	0	    &	4.49(-5)	&	4.01(-5)	&	0.05	\\
$p-$H$_2$O	$3_{22}-3_{13}$	&	1919.3595	&	204.7	&	2.25(-5)	&	1.50(-5)	&	0.18	\\
$p-$H$_2$O	$3_{13}-2_{02}$	&	2164.132	&	100.8	&	7.57(-5)	&	1.10(-4)	&	-0.16	\\
$p-$H$_2$O	$4_{04}-3_{13}$	&	2391.5726	&	204.7	&	3.08(-5)	&	5.15(-5)	&	-0.22	\\
$p-$H$_2$O	$3_{22}-2_{11}$	&	3331.4584	&	136.9	&	1.13(-4)	&	4.00(-5)	&	0.45	\\   
  \hline
\end{tabular}
\end{center}
Note: Entries in the table are: transition, Frequency (GHz), lower
level energy (K), observed and isobaric model line fluxes in CGS units
used by the Meudon PDR code (erg\,s$^{-1}$\,cm$^{-2}$\,sr$^{-1}$), and
the fractional difference between observations and the model. 
\end{table*}

\begin{figure}
\centering
\includegraphics[trim=1.5cm 2.3cm 6.0cm 2.0cm, clip=true, width=0.95\columnwidth]{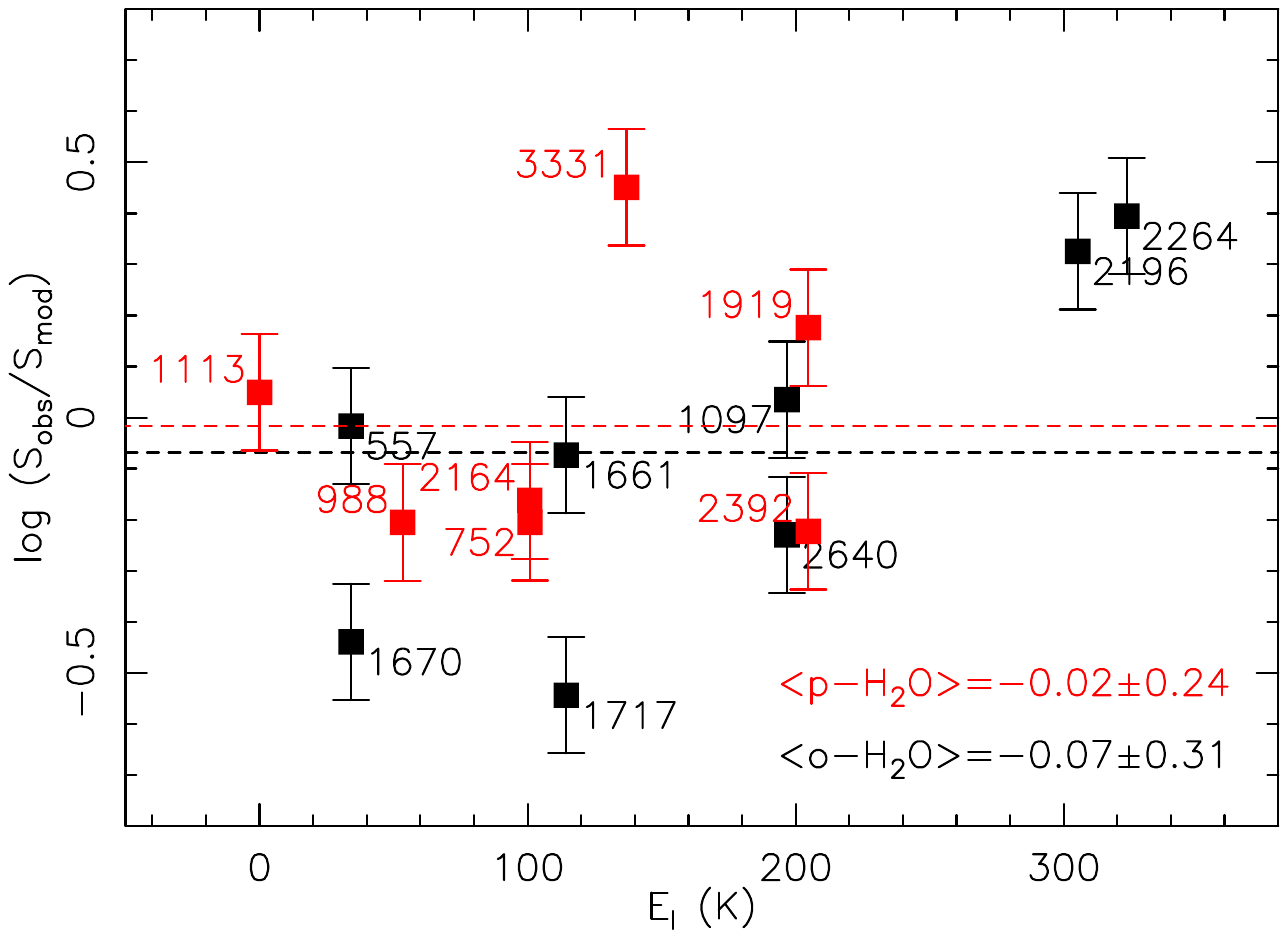} 
\caption{A comparison of the observed water line fluxes toward
  S140~IRS1 with the predictions of the best-fit isobaric PDR model.
  The logarithm of the observed to modeled flux ratio is
  plotted as a function of the lower level energy of the transition in
  K. Ortho- and para-water lines are plotted as black squares and red
  triangles, respectively. Average values are marked with the dashed
  lines. The points are labeled by the line frequency in GHz.
  Observational error bars of 30\% are assumed.}\label{fig:water}
\end{figure}

Figure~\ref{fig:water} shows the logarithm of the observed
to modeled line flux ratio, $\log (S_{o}/S_{m})$, plotted as a
function of the lower level energy of the transition in K. While the
scatter is significant, no significant trends can be seen. The mean
ratios between the observed and model line fluxes are $-0.11 \pm 0.22$
and $-0.02 \pm 0.24$ for the ortho- and para-water lines,
respectively. On average, our preferred isobaric model thus predicts
the water line fluxes consistent with observations. Some parameters of
the ISMDB models are generic, such as the physical properties and size
distribution of dust grains, the cosmic-ray ionization rate, and the
elemental abundances. These parameters could be optimized to better
match the S140 models; however, this is beyond the scope of the
present paper. We thus conclude that our simple plane-parallel PDR
model provides a satisfactory fit to the water observations and we see
no evidence for departures from the LTE water ortho-para ratio at the
local gas temperature, which is assumed in the model. This conclusion
is consistent with the published analysis of the Orion Bar water
observations \citep{Putaud2019}.

\begin{figure}
\centering
\includegraphics[trim=2.5cm 3cm 2.4cm 6cm, clip=true, width=0.95\columnwidth]{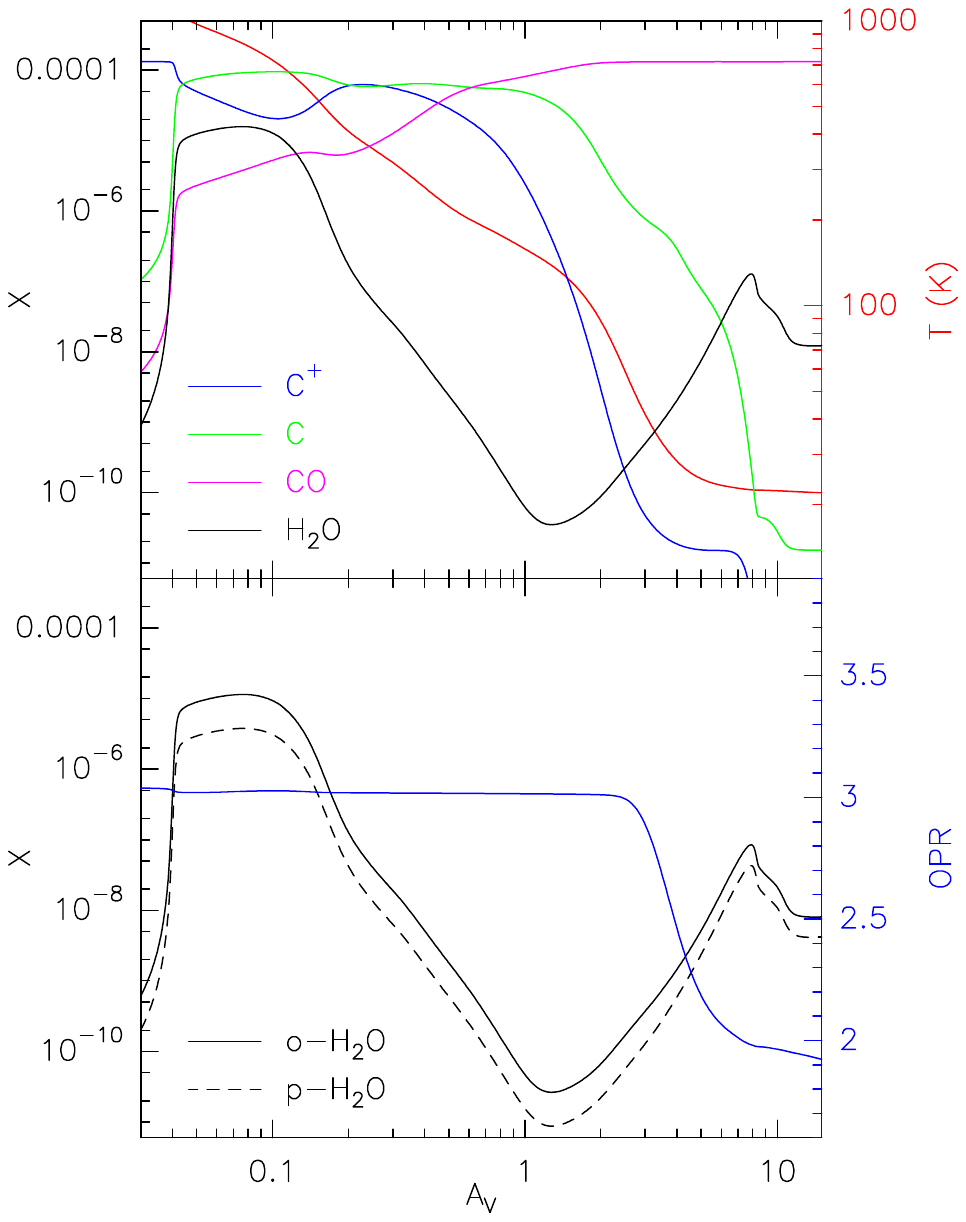} 
\caption{Abundance profiles with respect to hydrogen nuclei of C$^+$,
  C$^0$, CO, and H$_2$O plotted as a function of $A_V$ for the best
  fit isobaric model of S140~IRS1. The water abundance peaks
  near the surface of the PDR, with a secondary peak present in a
  shielded region at $A_V \sim 9$, where the gas temperature is
  $\sim 30$~K (red curve, right scale). The peak H$_2$O abundance in
  this region is $\sim 1.4 \times 10^{-7}$. }\label{fig:abundances}
\end{figure}

Figure~\ref{fig:abundances} (upper panel) shows abundance
profiles of the main gas-phase species in our preferred isobaric PDR
model. The water abundance (dark blue line) has two peaks: a narrow
one in the outer PDR layers where the high-temperature gas-phase
chemistry channel is active \citep{vanDishoeck2021} and a broader one
in a shielded region deeper into the cloud. This is similar to the
Orion Bar PDR results of \cite{Putaud2019}. 
Figure~\ref{fig:abundances} (lower panel) shows abundances of ortho-
and para-water (black curves, left scale) and the resulting OPR (blue
curve, right scale) as a function of $A_V$. The OPR drops from the
high temperature limit of 3 in the outer layers to $\sim 2$ in the
shielded interior, as expected given the temperature profile (red
line, right scale in the upper panel). Our PDR modeling approach
takes into account the temperature structure and the optical depth of
the lines. It thus offers a better modeling framework compared to
single slab models used in some previous studies of similar regions.
Optically thick ground-state water lines trace primarily the outer
abundance peak near the PDR surface \citep{Lis2026}.

The peak H$_2$O abundance near the PDR surface is
$\sim 1.6 \times 10^{-5}$ and in the shielded region is
$\sim 1.2 \times 10^{-7}$ at an $A_V$ of $\sim 8$, close to the peak
Orion Bar value of $\sim 2 \times 10^{-7}$ at an $A_V$ of $\sim 9$
\citep{Putaud2019}. For comparison, PDR models of \cite{Lis2026}
suggest that the water abundance in the low-UV illumination Horsehead
nebula PDR reaches $3.6 \times 10^{-7}$ at an $A_V \sim 3$. We thus
see no evidence for strong variations in the gas-phase water abundance
in PDRs as a function of the UV field strength, in agreement with the
theoretical predictions \citep{Hollenbach2009}.

\section{Summary and conclusions}
\label{sec:summary}

We analyzed archival \emph{Herschel} PACS, SPIRE FTS, and HIFI
spectroscopic observations of the S140 photon dominated region to
constrain the water abundance and ortho-para ratio in this region. The
main results of this study can be summarized as follows:

\begin{enumerate}

\item Toward the high-mass protostar IRS1, 52 spectral lines,
  including 15 water lines, are detected with the PACS and SPIRE
  instruments. HIFI strips of CO~9--8, $^{13}$CO 10--9, and
  H$_2$O~$1_{10}-1_{01}$ and $1_{11}-0_{00}$ lines show small spatial
  stratification between different tracers compared to nearby edge-on
  PDRs, which is likely related to the complex morphology of
  the source.

\item Best-fit PDR model including all CO lines detected toward
  IRS1 fails to accurately reproduce the shape of the CO spectral line
  energy distribution toward IRS1, over-predicting fluxes of the
  mid-$J$ CO lines, between $J=14-13$ and $23-22$ and under-predicting
  fluxes of the high-$J$ CO lines $J=28-27$ and above. This may
  suggest that processes other than UV illumination are at play in
  this region.

\item The CO rotation diagram toward S140 IRS1 is well described by
  two components with rotational temperatures of $133 \pm 19$ and
  $938 \pm 102$~K, respectively. This is consistent with
  \emph{Herschel} studies of similar sources and further indicates the
  presence of the hot shocked component along the line of sight. The
  spectral resolution of the PACS and SPIRE spectrometers is
  insufficient to separate the PDR and shock contributions to the
  observed line fluxes.

\item Best-fit isobaric PDR model of IRS1 excluding CO lines
  $J=15-14$ and above has a UV field enhancement factor
  $G_0=1 \times 10^3$ and Habing units and a pressure
  $p_{th}=5 \times 10^8$~K\,cm$^{-3}$. The derived pressure is
  comparable to that derived for the high-UV illumination PDR, the
  Orion Bar, while $G_0$ is an order of magnitude lower.

\item PDR models suggest that optically thick ground-state water lines
  toward IRS1 originate from the surface of the PDR. The gas
  temperature in this region is predicted to be $\sim 800$~K and the
  water abundance with respect to hydrogen nuclei reaches
  $\sim 10^{-5}$. However, bulk of the gas-phase water is
  located in a more shielded region with a gas temperature
  $\sim 24$~K, where gas-phase H$_2$O abundance reaches
  $\sim 1 \times 10^{-7}$.

\item A comparison of the observed water
  line intensities with the PDR model predictions shows no evidence
  for departures from the LTE ortho-para ratio at the local gas
  temperature, which is assumed in the models. These results are
  consistent with the Orion Bar analysis of \cite{Putaud2019}. We see
  no evidence for very low OPR values suggested in some earlier
  studies of UV-illuminated ISM regions.

\item The PDR model succeeds in reproducing the CO and [C\,{\sc i}]
  line intensities toward the externally illuminated PDR on the
  surface of the molecular cloud, where outflows and shocks do not
  contribute to the observed line emission. The best-fit isobaric
  model yields a pressure $p = 1 \times 10^6$~K\,cm$^{-3}$ and
  $G_0 = 2 \times 10^4$. However, the sensitivity of PACS and SPIRE
  was insufficient to detect water lines toward this region.
  
\end{enumerate}

The spectroscopic sensitivity of the \emph{Herschel} instruments
limited measurements of the water ortho-para ratio to bright PDRs in
active high-mass star forming regions, such as the Orion Bar or
S140~IRS1. However, future far-infrared space observatories, such as
the NASA Astrophysics Probe, PRIMA, currently in Phase A, will extend
such measurements to lower-UV illumination environments. This will
allow to determine how useful OPR measurements are for understanding
the origins of Solar System materials.

\begin{acknowledgements}
  
  This research was carried out at the Jet Propulsion Laboratory,
  California Institute of Technology, under a contract with the
  National Aeronautics and Space Administration (80NM0018D0004).
  D.C.L. and S.L. acknowledge financial support from the National
  Aeronautics and Space Administration (NASA) Astrophysics Data
  Analysis Program (ADAP). J.R.G. thanks the Spanish MICIU for funding
  support under grant PID2023-146667NB-I00. This work was supported by
  the Thematic Action “Physique et Chimie du Milieu Interstellaire”
  (PCMI) of INSU Programme National “Astro”, with contributions from
  CNRS Physique \& CNRS Chimie, CEA, and CNES.

\end{acknowledgements}

\appendix       

\section{\emph{Herschel} observations of S140} 
\label{app:obs}

The SPIRE spectrometer included two bolometer arrays with overlapping
bands of $194 - 313$~\mic\ (SSW) and $303-671$~\mic\ (SLW). The
highest unapodised resolution available was 0.04~cm$^{-1}$ (1.2~GHz),
corresponding to $R=\lambda/\Delta\lambda$ between 1300 at the
shortest wavelengths and 370 at the longest wavelengths. The two
hexagonal spectrometer arrays contained 37 and 19 detectors for SSW
and SLW, respectively (see instrument footprint in
Figure~\ref{fig:Spitzer}). The instrument FWHM beam size varied
between 17 and 42\arcs. Reduced and calibrated data products were
downloaded from the ESA \emph{Herschel} Science
Archive\footnote{https://archives.esac.esa.int/hsa/whsa/}. Spectra in
FITS format were then extracted into an ASCII file using a Python
script and imported into the IRAM CLASS software
package\footnote{https://www.iram.fr/IRAMFR/GILDAS/doc/html/class-html/class.html}
for baseline removal, Gaussian fitting, etc.

The SPIRE FTS and PACS line fluxes toward S140~ IRS1 are listed in
Tables~\ref{tab:spire} and \ref{tab:pacs}. The HIFI line intensities
of the ground-state ortho- and para-water lines along a strip through
S140~IRS1, perpendicular to the surface are listed in
Tables~\ref{tab:hifi}. Table~\ref{tab:spireif} lists the lines detected
with SPIRE~FTS at positions A and B toward the surface PDR front.

The PACS spectrometer had 25 spatial pixels (spaxels) with a spaxel
size of 9\farcs4 at all wavelengths. To ensure a better match with
the larger SPIRE FTS beam, we averaged the central $3\times3$ spaxels
corresponding to a $\sim 30 \times 30\arcs$ aperture. The resolving power
varied between 1500 and 2000 in the 55 -- 190~\mic\ wavelength range.
We used the re-binned data products (HPS3DRR/B) which have the native
irregular PACS footprint with 9\farcs4 spaxels. Spectra in FITS
format were extracted into an ASCII file using a Python script and
imported into the CLASS package.

To ensure that we correctly extracted and coadded the $3 \times 3$
central PACS spaxels, we compared similarly processed Orion Bar PACS
observations downloaded from the \emph{Herschel} archive with the
published observations of this source \citep{Joblin2018}. We found a
good agreement with the line fluxes from the literature, which
validates our data extraction procedure.

We used spectra obtained with the HIFI wideband spectrometer (WBS),
which we interpolated on a regular grid with a 10\arcs\ spacing in
right ascension, averaging the two instrumental polarizations. The
latest values of the HIFI beam efficiencies of 0.635 at 557~GHz and
0.59 at 1113~GHz were applied to convert the spectra to the main beam
brightness temperature scale. The HIFI beam size was $\sim 38$\arcs\
at 557~GHz and $\sim 19$\farcs5 at 1113~GHz. The velocity integrated line
intensities along the strip observed with HIFI are listed in
Table~\ref{tab:hifi}.

\begin{table}[ht]
\begin{center}  
\caption{Lines detected in the SPIRE FTS observation toward S140 IRS1.} 
\label{tab:spire}
\begin{tabular}{lccc}
\hline \hline 
  \rule[-3mm]{0mm}{8mm}  Line & $\nu$ & $S_L$ & $\delta S_L$  \\

  \hline 
  CO 4--3   &  461.0408	&  3.05(-5)	&	4.90(-7)   \\
  CO 5--4   &  576.2679	&  5.48(-5)	&	7.24(-7)   \\
  CO 6--5   &  691.4731	&  9.39(-5)	&	5.93(-7)   \\
  CO 7--6   &  806.6518	&  1.20(-4)	&	1.70(-6)   \\ 
  CO 8--7   &  921.7997	&  1.69(-4)	&	1.73(-6)   \\
  CO 9--8   & 1036.9124	&  2.44(-4)	&	3.53(-6)   \\ 
  CO 10--9  & 1151.9855	&  3.23(-4)	&	5.50(-6)   \\
  CO 11--10 & 1267.0145	&  3.49(-4)	&	6.73(-6)   \\
  CO 12--11 & 1381.9951	&  3.32(-4)	&	1.44(-5)   \\ 
  CO 13--12 & 1496.9229	&  4.04(-4)	&	3.49(-5)   \\
  \hline
  $^{13}$CO 5--4   &  550.9263  &	1.68(-5)	&	6.01(-7)  \\
  $^{13}$CO 6--5   &  661.0673  &	2.05(-5)	&	5.58(-7)  \\
  $^{13}$CO 7--6   &  771.1841  &	3.32(-5)	&	9.99(-7)  \\
  $^{13}$CO 8--7   &  881.2728  &	2.22(-5)	&	1.26(-6)  \\
  $^{13}$CO 9--8   &  991.3293  &	2.49(-5)	&	1.98(-6)  \\
  $^{13}$CO 10--9  & 1101.3497  &	2.73(-5)	&	5.62(-6)  \\
  \hline
  H$_2$O $1_{10}-1_{01}$  &  556.9360  &	6.68(-6)  &	4.97(-7)  \\
  H$_2$O $2_{11}-2_{02}$  &  752.0331  &	1.19(-5)  &	8.00(-7)  \\
  H$_2$O $2_{02}-1_{11}$  &  987.9268  &	3.98(-5)  &	3.09(-6)  \\
  H$_2$O $3_{12}-3_{03}$  & 1097.3648  &	3.48(-5)  &	5.81(-6)  \\
  H$_2$O $1_{11}-0_{00}$  & 1113.3430  &	4.49(-5)  &	5.71(-6)  \\
  \hline
  \CI\ $^3P_1$--$^3P_0$    &  492.1607 &	6.54(-6)  &	5.21(-7)  \\
  \CI\ $^3P_2$--$^3P_1$    &  809.3418 &	1.66(-5)  &	1.74(-6)  \\
  \NII\ $^3P_1 $--$^3P_0$  & 1461.1314  &	1.37(-4)  &	1.65(-5)  \\
  HCO$^+$ 7--6         &  624.2084  &	1.33(-5)  &	5.88(-7)  \\
  HCO$^+$ 8--7         &  713.3412  &	5.46(-6)  &	5.02(-7)  \\
  \hline
\end{tabular}
\end{center}
Note: Entries in the table are: transition, Frequency (GHz) line flux
and the corresponding uncertainty from the Gaussian fit in CGS units
used by the Meudon PDR code (erg\,s$^{-1}$\,cm$^{-2}$\,sr$^{-1}$). The
line frequency was fixed in the Gaussian fit to the H$_2$O
$1_{10}-1_{01}$ line, and the line width was also fixed to that of the
nearby stronger CO and $^{13}$CO lines.
\end{table}

\begin{table}[ht]
\begin{center}  
\caption{Lines detected in the PACS observation toward S140 IRS1.} %
\label{tab:pacs}
\begin{tabular}{lccc}
\hline \hline 
  \rule[-3mm]{0mm}{8mm}  Line & $\nu$ & $S_L$ & $\delta S_L$  \\

  \hline 
  CO 14--13	& 1611.7935  &	2.95(-4)	&	4.49(-6) \\
  CO 15--14	& 1726.6025  &	2.47(-4)	&	3.42(-6) \\
  CO 16--15	& 1841.3455  &	2.15(-4)	&	3.01(-6) \\
  CO 17--16	& 1956.0181  &	1.80(-4)	&	3.46(-6) \\
  CO 18--17	& 2070.6160  &	1.87(-4)	&	4.08(-6) \\
  CO 19--18	& 2185.1347  &	1.71(-4)	&	9.83(-6) \\
  CO 20--19	& 2299.5698  &	1.89(-4)	&	4.24(-6) \\
  CO 21--20	& 2413.9171  &	1.66(-4)	&	7.49(-6) \\
  CO 22--21	& 2528.1721  &	1.10(-4)	&	6.12(-6) \\
  CO 23--22	& 2642.3303  &	9.30(-5)	&	1.56(-5) \\
  CO 24--23	& 2756.3876  &	8.72(-5)	&	7.92(-6) \\
  CO 28--27	& 3211.5187  &	7.77(-5)	&	9.06(-6) \\
  CO 29--28	& 3325.0053  &	6.65(-5)	&	8.52(-6) \\
  CO 30--29	& 3438.3646  &	1.06(-4)	&	1.50(-5) \\
  \hline
  H$_2$O $2_{21}-2_{12}$	& 1661.0076  &	5.58(-5)	&	3.99(-6) \\
  H$_2$O $2_{12}-1_{01}$	& 1669.9048  &	6.25(-5)	&	3.57(-6) \\
  H$_2$O $3_{03}-2_{12}$	& 1716.7696  &	6.11(-5)	&	3.31(-6) \\
  H$_2$O $3_{22}-3_{13}$	& 1919.3595  &	2.25(-5)	&	6.16(-6) \\
  H$_2$O $3_{13}-2_{02}$	& 2164.1320  &	7.57(-5)	&	3.69(-6) \\
  H$_2$O $3_{30}-3_{21}$	& 2196.3458  &	2.37(-5)	&	3.02(-6) \\
  H$_2$O $4_{23}-4_{14}$	& 2264.1497  &	3.30(-5)	&	4.30(-6) \\
  H$_2$O $4_{04}-3_{13}$	& 2391.5726  &	3.08(-5)	&	3.75(-6) \\
  H$_2$O $4_{14}-3_{03}$	& 2640.4738  &	1.09(-4)	&	1.06(-5) \\
  H$_2$O $3_{22}-2_{11}$	& 3331.4584  &	1.13(-4)	&	1.12(-5) \\
  \hline
  \CII\ $^3P_{3/2}$--$^3P_{1/2}$	& 1900.5369  &	5.70(-5)	&	5.11(-6)) \\
  \OI\ $^3P_0$--$^3P_1$	& 2060.0689  &	4.50(-4)	&	3.85(-6)) \\
  \hline
\end{tabular}
\end{center}
Note: Entries in the table are: transition, Frequency (GHz) line flux
and the corresponding uncertainty from a Gaussian fit in CGS units
used by the Meudon PDR code (erg\,s$^{-1}$\,cm$^{-2}$\,sr$^{-1}$).
\vspace{10mm}
\end{table}

\begin{table}[htb!]
\begin{center}  
\caption{Lines detected in the SPIRE FTS observation toward S140
  surface PDR.} 
\label{tab:spireif}
\begin{tabular}{lccccccc}
\hline \hline 
  \rule[-3mm]{0mm}{8mm}  Line & $\nu$ & $S_L$(A) & $\delta S_L$(A) & $S_L$(B) & $\delta S_L$(B) & $S_L$(Ave) & $\delta S_L$(Ave)  \\

  \hline 
  CO 4--3   &  461.0408	&  1.20(-5)	&	3.59(-7)  &  1.09(-5)	&	3.41(-7) &  1.15(-5)	&	3.50(-7)\\
  CO 5--4   &  576.2679	&  2.11(-5)	&	5.45(-7)  &  1.88(-5)	&	4.15(-7) &  2.00(-5)	&	4.16(-7) \\
  CO 6--5   &  691.4731	&  2.53(-5)	&	2.67(-7)  &  2.45(-5)	&	3.41(-7) &  2.49(-5)	&	3.11(-7) \\
  CO 7--6   &  806.6518	&  2.74(-5)	&	8.35(-7)  &  2.61(-5)	&	1.22(-6) &  2.68(-5)	&	9.67(-7) \\ 
  CO 8--7   &  921.7997	&  2.84(-5)	&	7.49(-7)  &  3.28(-5)	&	1.18(-6) &  3.06(-5)	&	9.65(-7) \\
  CO 9--8   & 1036.9124	&  1.52(-5)	&	8.89(-7)  &  2.12(-5)	&	1.28(-6) &  1.82(-5)	&	1.08(-6) \\ 
  CO 10--9  & 1151.9855	&  1.69(-5)	&	1.10(-6)  &  1.83(-5)	&	1.25(-6) &  1.76(-5)	&	1.18(-6) \\
  \hline 
  $^{13}$CO 5--4   &  550.9263	&  5.23(-6)	&	3.11(-7)  &  4.67(-6)	&	3.62(-6) &  4.95(-6)	&	3.37(-7) \\
  $^{13}$CO 6--5   &  661.0673	&  5.72(-6)	&	2.30(-7)  &  6.54(-6)	&	2.54(-6) &  6.13(-6)	&	2.42(-7) \\
  $^{13}$CO 7--6   &  771.1841	&  2.73(-5)	&	5.25(-7)  &  2.60(-5)	&	2.60(-5) &  2.67(-5)	&	7.63(-7) \\
  $^{13}$CO 8--7   &  881.2728	&  2.20(-6)	&	5.71(-7)  &  --	& -- &  --	&	-- \\ 
  \hline 
  \CI\ $^3P_1$--$^3P_0$    &  492.1607 &	3.69(-6)  &	3.66(-7) &  --	&	-- &  --	&	-- \\
  \CI\ $^3P_2$--$^3P_1$    &  809.3418 &	8.98(-6)  &	7.33(-7) &  9.82(-6)	&	1.29(-6) &  9.40(-6)	&	1.01(-6) \\
  \hline 
  \CII\ $^3P_{3/2}$--$^3P_{1/2}$	& 1900.5369  & & & & &	6.82(-4)	&	4.2(-6) \\
  \hline
\end{tabular}
\end{center}
Note: Entries in the table are: transition, Frequency (GHz) line flux
and the corresponding uncertainty from a Gaussian fit in CGS units
used by the Meudon PDR code (erg\,s$^{-1}$\,cm$^{-2}$\,sr$^{-1}$). The 
last two columns correspond to an average of positions A and B. The
[C\,{\sc ii}] line flux is from \cite{Dedes2010}.
\end{table}

\begin{table}[htb!]
\begin{center}  
\caption{Lines intensities along the strip observed with HIFI.} %
\label{tab:hifi}
\begin{tabular}{rrrrrr}
\hline \hline 
  \rule[-3mm]{0mm}{8mm} $\Delta\alpha$ & $\Delta\delta$ & $I$(CO) &
       $I$($^{13}$CO) & $I$($o-$H$_2$O) & $I$($p-$H$_2$O) \\
  \hline 
-65	&	-86	&	1.65	&	1.00	&	0.34	&	0.28	\\
-60	&	-80	&	3.12	&	0.20	&	0.30	&	-0.55	\\
-55	&	-73	&	3.10	&	0.28	&	0.24	&	-0.63	\\
-50	&	-66	&	4.64	&	0.23	&	0.30	&	-0.50	\\
-45	&	-60	&	12.99	&	0.64	&	0.88	&	-0.19	\\
-40	&	-53	&	22.53	&	0.46	&	1.44	&	0.01	\\
-35	&	-46	&	32.41	&	-0.71	&	0.96	&	1.75	\\
-30	&	-40	&	43.69	&	0.58	&	2.56	&	0.76	\\
-25	&	-33	&	70.53	&	1.59	&	3.75	&	2.34	\\
-20	&	-26	&	122.0	&	2.96	&	5.32	&	4.07	\\
-15	&	-20	&	181.8	&	9.23	&	8.30	&	8.21	\\
-10	&	-13	&	269.2	&	21.56	&	11.89	&	15.34	\\
-5	&	-7	&	348.5	&	38.49	&	15.96	&	21.75	\\
0	&	0	&	344.2	&	42.27	&	18.88	&	19.18	\\
5	&	7	&	267.2	&	31.27	&	19.56	&	13.16	\\
10	&	13	&	166.5	&	15.91	&	18.04	&	8.36	\\
15	&	20	&	95.61	&	4.91	&	15.24	&	6.14	\\
20	&	27	&	57.92	&	1.92	&	11.62	&	4.28	\\
25	&	33	&	40.55	&	0.31	&	8.30	&	3.23	\\
30	&	40	&	29.06	&	0.33	&	6.22	&	1.82	\\
35	&	47	&	21.13	&	0.03	&	5.08	&	1.97	\\
40	&	53	&	13.56	&	1.35	&	3.53	&	0.36	\\
45	&	60	&	10.83	&	0.39	&	2.48	&	0.45	\\
50	&	67	&	7.32	&	0.01	&	2.28	&	0.89	\\
55	&	73	&	5.53	&	0.81	&	1.99	&	0.41	\\
60	&	80	&	2.53	&	0.59	&	1.14	&	-0.01	\\
65	&	87	&	3.27	&	1.03	&	0.92	&	-0.18	\\            
  \hline
\end{tabular}
\end{center}
Note: Entries in the table are: right ascension and declination
offsets from the reference position (\arcs), line intensities of
CO~9--8, $^{13}$CO~10--9, $o-$H$_2$O~$1_{10}-1_{01}$, and
$p-$H$_2$O~$1_{11}-0_{00}$ integrated between --12~km\,s$^{-1}$ and 0~km\,s$^{-1}$
(K\,km\,s$^{-1}$). 
\end{table}

\end{document}